\documentclass[12pt,twoside,a4paper]{article}
\usepackage[utf8]{inputenc}
\usepackage[margin=2cm]{geometry}
\usepackage{tikz}
\usepackage{animate}
\usepackage{amsmath,amsthm,mathtools,amssymb,lscape}
\usepackage{thmtools,thm-restate}
\usepackage{mathrsfs}
\usepackage{proba}
\usepackage{booktabs}
\usepackage{threeparttable,subfigure}
\usepackage{enumerate}
\usepackage[shortlabels]{enumitem}
\usepackage[round]{natbib}
\usepackage{graphicx,setspace,multirow}
\usepackage{epsfig,tabularx, subfigure,qtree,pict2e}
\usepackage{dsfont}
\usepackage{caption}
\usepackage{xcolor}
\usepackage{chngcntr}

\usepackage{hyperref}
\hypersetup{
    unicode=false,          % non-Latin characters in Acrobat’s bookmarks
    pdftoolbar=true,        % show Acrobat’s toolbar?
    pdfmenubar=true,        % show Acrobat’s menu?
    pdffitwindow=false,     % window fit to page when opened
    pdftitle={},    % title
    pdfauthor={},     % author
    pdfsubject={},   % subject of the document
    pdfcreator={},   % creator of the document
    pdfproducer={},  % producer of the document
    pdfkeywords={}, % list of keywords
    pdfnewwindow=true,      % links in new window
    colorlinks=false,       % false: boxed links; true: colored links
    linkcolor=red,          % color of internal links
    citecolor=green,        % color of links to bibliography
    filecolor=magenta,      % color of file links
    urlcolor=cyan           % color of external links
}

\newtheorem{Def}{Definition}
\newtheorem{algorithm}{Algorithm}
\newtheorem{assumption}{Assumption}

\numberwithin{equation}{section}

\def\wl{\par \vspace{\baselineskip}}
\def\m{\mathcal}
\def\b{\boldsymbol}
\def\vec{\mathsf{vec}\,}

\def\E{\mathbb{E}}
\def\R{\mathbb{R}}

\def\MSPE{\mathsf{MSPE}}
\def\RMSPE{\mathsf{RMSPE}}
\def\MAPE{\mathsf{MAPE}}
\def\cp{\overset{p}{\longrightarrow}}

\def\1{\mathds{1}}

\DeclareSymbolFont{sfoperators}{OT1}{cmss}{m}{n}
\DeclareSymbolFontAlphabet{\mathsf}{sfoperators}

\makeatletter
\def\operator@font{\mathgroup\symsfoperators}
\makeatother

\begin{document}

\title{Machine Learning Advances for Time Series Forecasting}

\author{
{\textbf{Ricardo P. Masini}} \\
\small{São Paulo School of Economics, Getulio Vargas Foundation} \\
\small{E-mail: \texttt{ricardo.masini@fgv.br}
\vspace{0.3cm}}
\and
{\textbf{Marcelo C. Medeiros}} \\
\small{Department of Economics, Pontifical Catholic University of Rio de Janeiro} \\
\small{E-mail: \texttt{mcm@econ.puc-rio.br}
\vspace{0.3cm}}\and
{\textbf{Eduardo F. Mendes}} \\ \small{School of Applied Mathematics, Getulio Vargas Foundation} \\
\small{E-mail: \texttt{eduardo.mendes@fgv.br}\vspace{0.3cm}}}

\maketitle

\vspace{0.1cm}

\begin{abstract}
\noindent
In this paper we survey the most recent advances in supervised machine learning and high-dimensional models for time series forecasting. We consider both linear and nonlinear alternatives. Among the linear methods we pay special attention to penalized regressions and ensemble of models. The nonlinear methods considered in the paper include shallow and deep neural networks, in their feed-forward and recurrent versions, and tree-based methods, such as random forests and boosted trees. We also consider ensemble and hybrid models by combining ingredients from different alternatives. Tests for superior predictive ability are briefly reviewed. Finally, we discuss application of machine learning in economics and finance and provide an illustration with high-frequency financial data.
\wl
\wl
\wl
\noindent
\textbf{Acknowledgements}: We are very grateful for the insightful comments made by two anonymous referees. The second author gratefully acknowledges the partial financial support from CNPq. We are also grateful to Francis X. Diebold, Daniel Borup, and Andrii Babii for helpful comments.

\end{abstract}

\onehalfspacing

\newpage

\section{Introduction}
This paper surveys the recent developments in Machine Learning (ML) methods to economic and financial time series forecasting. ML methods have become an important estimation, model selection and forecasting tool for applied researchers in Economics and Finance. With the availability of vast datasets in the era of \emph{Big Data}, producing reliable and robust forecasts is of great importance.\footnote{More recently, ML for causal inference have started to receive a lot of attention. However, this survey will not cover causal inference with ML methods.}

However, what is Machine Learning? It is certainly a buzzword which has gained a lot of popularity during the last few years. There are a myriad of definitions in the literature and one of the most well established is from the artificial intelligence pioneer Arthur L. Samuel who defines ML as the ``the field of study that gives computers the ability to learn without being explicitly programmed.''\footnote{The original sentence is ``Programming computers to learn from experience should eventually eliminate the need for much of this detailed programming effort.'' See, \citet{aS1959}.} We prefer a less vague definition where ML is the combination of automated computer algorithms with powerful statistical methods to learn (discover) hidden patterns in rich datasets. In that sense, \emph{Statistical Learning Theory} gives the statistical foundation of ML. Therefore, this paper is about Statistical Learning developments and not ML in general as we are going to focus on statistical models. ML methods can be divided into three major groups: supervised, unsupervised, and reinforcement learning. This survey is about supervised learning, where the task is to learn a function that maps an input (explanatory variables) to an output (dependent variable) based on data organized as input-output pairs. Regression models, for example, belongs to this class. On the other hand, unsupervised learning is a class of ML methods that uncover undetected patterns in a data set with no pre-existing labels as, for example, cluster analysis or data compression algorithms. Finally, in reinforcement learning, an agent learns to perform certain actions in an environment which lead it to maximum reward. It does so by exploration and exploitation of knowledge it learns by repeated trials of maximizing the reward. This is the core of several artificial intelligence game players (AlfaGo, for instance) as well as in sequential treatments, like Bandit problems.

The  supervised ML methods presented here can be roughly divided in two groups. The first one includes linear models and are discussed in Section \ref{S:Linear}. We focus mainly on specifications estimated by regularization, also known as shrinkage. Such methods date back at least to \citet{anT1943}. In Statistics and Econometrics, regularized estimators gained attention after the seminal papers by Willard James and Charles Stein who popularized the bias-variance trade-off in statistical estimation \citep{cS1956,wJcS1961}. We start by considering the Ridge Regression estimator put forward by \citet{aHrK1970}. After that, we present the Least Absolute Shrinkage and Selection (LASSO) estimator of \citet{rT1996} and its many extensions. We also include a discussion of other penalties. Theoretical derivations and inference for dependent data are also reviewed.

The second group of ML techniques focus on nonlinear models. We cover this topic in Section \ref{S:nonlinear} and start by presenting an unified framework based on sieve semiparametric approximation as in \citet{uG1981}. We continue by analysing specific models as special cases of our general setup. More specifically, we cover feedforward neural networks, both in their shallow and deep versions and recurrent neural networks, and tree-based models such as random forests and boosted trees. Neural Networks (NN) are probably one of the most popular ML methods. The success is partly due to the, in our opinion, misguided analogy to the functioning of the human brain. Contrary of what has been boasted in the early literature, the empirical success of NN models comes from a mathematical fact that a linear combination of sufficiently many simple basis functions is able to approximate very complicated functions arbitrarily well in some specific choice of metric. Regression trees only achieved popularity after the development of algorithms to attenuate the instability of the estimated models. Algorithms like Random Forests and Boosted Trees are now in the toolbox of applied economists.

In addition to the models mentioned above, we also include a survey on ensemble-based methods such as Bagging \citet{lB1996} and the Complete Subset Regression \citep{gEaGaT2013,gEaGaT2015}. Furthermore, we give a brief introduction to what we named ``hybrid methods'', where ideas from both linear and nonlinear models are combined to generate new ML forecasting methods.

Before presenting an empirical illustration of the methods, we discuss tests of superior predictive ability in the context of ML methods.

\subsection{General Framework}\label{S:framework}
A quick word on notation: an uppercase letter as in $X$ denotes a random quantity as opposed to a lowercase letter $x$ which denotes a deterministic (non-random) quantity. Bold letters as in $\b X$ and $\b x$ are reserved for multivariate objects such as vector and matrices. The symbol $\|\cdot\|_q$ for $q\geq 1$ denotes the $\ell_q$ norm of a vector. For a set $S$ we use $|S|$ to denote its cardinality.

Given a sample with $T$ realizations of the random vector $\left(Y_t,\b{Z}_t'\right)'$, the goal is to predict $Y_{T+h} $ for horizons $h=1,\ldots,H$. Throughout the paper, we consider the following assumption:
\begin{assumption}[DGP]\label{A:DGP}
Let $\{(Y_t,\b{Z}_t')'\}_{t=1}^{\infty}$ be a covariance-stationary stochastic process taking values on $\R^{d+1}$.
\end{assumption}
Therefore, we are excluding important non-stationary processes that usually appear in time-series applications. In particular unit-root and some types on long-memory process are excluded by Assumption \ref{A:DGP}.

For (usually predetermined) integers $p\geq 1$ and $r\geq 0$ define the $n$-dimensional vector of predictors $\b{X}_t:=\left(Y_{t-1},\ldots,Y_{t-p},\b{Z}_t',\ldots,\b{Z}_{t-r}'\right)'$ where $n=p+d(r+1)$ and consider the following direct forecasting model:
\begin{equation}\label{eq:model}
Y_{t+h} = f_h(\b{X}_t) + U_{t+h},\quad h=1,\ldots,H,\quad t=1,\ldots,T,
\end{equation}
where $f_h:\R^n\to\R$ is an unknown (measurable) function and $U_{t+h}:= Y_{t+h} - f_h(\b X_t)$ is assumed to be zero mean and finite variance\footnote{The zero mean condition can be always ensured by including an intercept in the model. Also the variance of $f(\b X_t)$ to be finite suffices for the finite variance }.

The model $f_h$ could be the conditional expectation function, $f_h(\b x) = \E(Y_{t+h}|\b X_t=\ \b x)$, or simply the best linear projection of $Y_{t+h}$ onto the space spanned by $\b X_t$. Regardless of the model choice, our target becomes $f_h$, for $h=1,\ldots,H$. As $f_h$ is unknown, it should be estimated from data. The target function $f_h$ can be a single model or an ensemble of different specifications and it can also change substantially for each forecasting horizon.

Given an estimate $\widehat{f}_h$ for $f_h$, the next step is to evaluate the forecasting method by estimating its prediction accuracy. Most measures of prediction accuracy derives from the random quantity $\Delta_h(\b X_t):=|\widehat{f}_h(\b X_t) - f_h(\b X_t)|$. For instance the term \emph{prediction consistency} refers to estimators such that $\Delta_h(\b X_t)\cp 0$ as $T\to \infty$ where the probability is taken to be unconditional; as opposed to  its \emph{conditional} counterpart which  is given by $\Delta_h(\b x_t)\cp 0$, where the probability law is conditional on $\b X_t= \b x_t$. Clearly, if the latter holds for (almost) every $\b x_t$ then the former holds by the law of iterated expectation.

Other measures of prediction accuracy can be derived from the $\mathcal{L}_q$ norm induced by either the unconditional probability law $\E|\Delta_h(\b X_t)|^q$ or the conditional one  $\E(|\Delta_h(\b X_t)|^q|\b X_t = \b x_t)$ for $q\geq 1$.  By far, the most used are the \emph{(conditional) mean absolutely prediction error} ($\MAPE$) when $q=1 $ and \emph{(conditional) mean squared prediction error} ($\MSPE$) when $q=2$ or the \emph{(conditional) root mean squared prediction error} ($\RMSPE$) which is simply the square root of $\MSPE$. Those  measures of prediction accuracy based on the  $\mathcal{L}_q$ norms are stronger than prediction consistency in the sense that the converge to zero  as sample since increases of any of those $(q\geq 1)$ implies prediction consistency by Markov's inequality.

This approach stems from casting economic forecasting as a decision problem. Under the choice of a loss function, the goal is to select $f_h$ from a family of candidate models that minimises the the expected predictive loss or risk. Given a estimate $\widehat{f}_h$ for $f_h$, the next step is to evaluate the forecasting method by estimating its risk. The most commonly used losses are the absolute error and squared error, corresponding to $\mathcal{L}_1$ and $\mathcal{L}_2$ risk functions, respectively. See \citet{cGmM2006} for references a detailed exposition of this topic, \cite{gEaT2008} for a discussion of the role of loss function in forecasting, and \cite{gEaT2016} for a more recent review.

\subsection{Summary of the Paper}
Apart form this brief introduction, the paper is organized as follows. Section \ref{S:Linear} reviews  penalized linear regression models. Nonlinear ML models are discussed in Section \ref{S:nonlinear}. Ensemble and hybrid methods are presented in Section \ref{S:Other}.  Section \ref{S:Comp} briefly discusses tests for superior predictive ability. An empirical application is presented in Section \ref{S:App}. Finally, we conclude and discuss some directions for future research in Section \ref{S:Conclusions}.

\section{Penalized Linear Models}\label{S:Linear}

We consider the family of linear models where $f(\b x)=\b\beta_{0}'\b x$ in \eqref{eq:model} for a vector of unknown parameters $\b\beta _{0}\in\R^n$. Notice that we drop the subscript $h$ for clarity. However, the model as well as the parameter $\b\beta_0$ have to be understood for particular value of the forecasting horizon $h$. These models contemplate a series of well-known specifications in time series analysis, such as predictive regressions, autoregressive models of order $p$,  $AR(p)$, autoregressive models with exogenous variables,  $ARX(p)$, autoregressive models with dynamic lags $ADL(p,r)$, among many others \citep{jH1994tsbook}. In particular, \eqref{eq:model} becomes
\begin{equation}\label{eq:modely}
Y_{t+h} = \b\beta_{0}'\b{X}_t + U_{t+h},\quad h=1,\ldots,H,\quad t=1,\ldots,T,
\end{equation}
where, under squared loss, $\b \beta_{0}$ is identified by the best linear projection of $Y_{t+h}$ onto $\b X_t$ which is well defined whenever $\b\Sigma:=\E(\b X_t \b X_t')$ is non-singular. In that case, $U_{t+h}$ is orthogonal to $\b X_t$ by construction and this property is exploited to derive estimation procedures such as the Ordinary Least Squares (OLS). However, when $n>T$ (and sometimes $n\gg T$) the OLS estimator is not unique as the sample counterpart of $\b\Sigma$ is rank deficient. In fact, we can completely overfit whenever $n\geq T$.

Penalized linear regression arises in the setting where the regression parameter is not uniquely defined. It is usually the case when $n$ is large, possibly larger than the number of observations $T$, and/or when covariates are highly correlated. The general idea is to restrict the solution of the OLS problem to a ball around the origin. It can be shown that, although biased, the restricted solution has smaller mean squared error, when compared to the unrestricted OLS \cite[Ch. 3 and  Ch. 6]{ESLII2009}.

In penalized regressions the estimator $\widehat{\b\beta}$ for the unknown parameter vector $\b\beta_{0}$ minimizes the Lagrangian form
\begin{equation}\label{E:shrinkage}
\begin{split}
Q(\b\beta)&=\sum_{t=1}^{T-h}\left(Y_{t+h} -\b\beta'\b X_t\right)^2+ p(\b\beta),\\
&=\|\b{Y}-\b{X}\b\beta\|_2^2+ p(\b\beta),
\end{split}
\end{equation}
where $\b Y:=(Y_{h+1},\dots Y_T)'$, $\b X:= (\b X_1,\dots \b X_{T-h})'$ and $p(\b\beta):=p(\b\beta;\lambda,\b\gamma,\b{Z})\geq 0$ is a penalty function that depends on a tuning parameter $\lambda\geq 0$, that controls the trade-off between the goodness of fit and the regularization term. If $\lambda=0$, we have an the classical unrestricted regression, since $p(\b\beta;0,\b\gamma,\b{X})=0$. The penalty function may also depend on a set of extra hyper-parameters $\b\gamma$, as well as on the data $\b X$. Naturally, the estimator $\widehat{\b\beta}$ also depends on the choice of $\lambda$ and $\b\gamma$. Different choices for the penalty functions were considered in the literature of penalized regression.

\subsubsection*{Ridge Regression}
The ridge regression was proposed by \cite{aHrK1970} as a way to fight highly correlated regressors and stabilize the solution of the linear regression problem. The idea was to introduce a small bias but, in turn, reduce the variance of the estimator. The ridge regression is also known as a particular case of Tikhonov Regularization \citep{anT1943,aT1963,aTvA1977book}, in which the scale matrix is diagonal with identical entries.

The ridge regression corresponds to penalizing the regression by the squared $\ell_2$ norm of the parameter vector, i.e., the penalty in \eqref{E:shrinkage} is given by
\begin{equation*}
 p(\b\beta)=\lambda\sum_{i=1}^n \beta_{i}^2 = \lambda\|\b\beta\|_2^2.
\end{equation*}

Ridge regression has the advantage of having an easy to compute analytic solution, where the coefficients associated with the least relevant predictors are shrunk towards zero, but never reaching exactly zero. Therefore, it cannot be used for selecting predictors, unless some truncation scheme is employed.

\subsubsection*{Least Absolute Shrinkage and Selection Operator (LASSO)}

The LASSO was proposed by \citet{rT1996} and \citet{sCdDmS2001} as a method to regularize and perform variable selection at the same time. LASSO is one of the most popular regularization methods and it is widely applied in data-rich environments where number of features $n$ is much larger than the number of the observations.

LASSO corresponds to penalizing the regression by the $\ell_1$ norm of the parameter vector, i.e., the penalty in \eqref{E:shrinkage} is given by
\begin{equation*}
 p(\b\beta)=\lambda\sum_{i=1}^n |\beta_{i}| = \lambda\|\b\beta\|_1.
\end{equation*}

The solution of the LASSO is efficiently calculated by coordinate descent algorithms \citep[Ch. 5]{tHrTmW2015book}. The $\ell_1$ penalty is the smallest convex $\ell_p$ penalty norm that yields \emph{sparse} solutions. We say the solution is \emph{sparse} if only a subset $k<n$ coefficients are non-zero. In other words, only a subset of variables is selected by the method. Hence, LASSO is most useful when the total number of regressors $n\gg T$ and it is not feasible to test combination or models.

Despite attractive properties, there are still limitations to the LASSO. A large number of alternative penalties have been proposed to keep its desired properties whilst overcoming its limitations.

\subsubsection*{Adaptive LASSO}

The adaptive LASSO (adaLASSO) was proposed by \cite{hZ2006} and aimed to improve the LASSO regression by introducing a weight parameter, coming from a first step OLS regression. It also has sparse solutions and efficient estimation algorithm, but enjoys the \emph{oracle property}, meaning that it has the same asymptotic distribution as the OLS conditional on knowing the variables that should enter the model.\footnote{The \emph{oracle property} was first described in \cite{jFrL2001} in the context of non-concave penalized estimation.}

The adaLASSO penalty consists in using a weighted $\ell_1$ penalty:
\begin{equation*}
p(\b\beta)=\lambda\sum_{i=1}^n \omega_i|\beta_{i}|,
\end{equation*}
where $\omega_i=|\beta_{i}^*|^{-1}$ and $\beta_{i}^*$ is the coefficient from the first-step estimation (any consistent estimator of $\b\beta_0$) AdaLASSO can deal with many more variables than observations. Using LASSO as the first-step estimator can be regarded as the two-step implementation of the local linear approximation in \cite{jFlXhZ2014} with a zero initial estimate.

\subsubsection*{Elastic net}

The elastic-net (ElNet) was proposed by \cite{hZtH2005} as a way of combining strengths of LASSO and ridge regression. While the $L_1$ part of the method performs variable selection, the $L_2$ part stabilizes the solution. This conclusion is even more accentuated when correlations among predictors become high. As a consequence, there is a significant improvement in prediction accuracy over the LASSO \citep{hZhZ2009}.

The elastic-net penalty is a convex combination of $\ell_1$ and $\ell_2$ penalties:
\begin{equation*}
p(\b\beta)=\lambda\left[\alpha\sum_{i=1}^n\beta_{i}^2+(1-\alpha)\sum_{i=1}^n|\beta_{i}|\right]=\lambda[\alpha\|\b\beta\|_2^2 + (1-\alpha)\|\b\beta\|_1],
\end{equation*}
where $\alpha \in [0,1]$. The elastic net has both the LASSO and ridge regression as special cases.

Just like in the LASSO regression, the solution to the elastic-net problem is efficiently calculated by coordinate descent algorithms. \citet{hZhZ2009} proposes the adaptive elastic net. The elastic-net and adaLASSO improve the LASSO in distinct directions: the adaLASSO has the oracle property and the elastic net helps with the correlation among predictors. The adaptive elastic-net combines the strengths of both methods. It is a combination of ridge and adaLASSO, where the first-step estimator come from the elastic-net.

\subsubsection*{Folded concave penalization}
 % SCAD and MCP
 LASSO approaches became popular in sparse high-dimensional estimation problems largely due their computational properties. Another very popular approach is the folded concave penalization of \cite{jFrL2001}. This approach covers a collection of penalty functions satisfying a set of properties. The penalties aim to penalize more parameters close to zero than those that are further away, improving performance of the method. In this way, penalties are concave with respect to each $|\beta_i|$.

 One of the most popular formulations is the SCAD (smoothly clipped absolute deviation). Note that unlike LASSO, the penalty may depend on $\lambda$ in a nonlinear way. We set the penalty in \eqref{E:shrinkage} as  $p(\b\beta) = \sum_{i=1}^n \widetilde{p}(\beta_i,\lambda,\gamma)$ where
\begin{equation*}
    \widetilde{p}(u,\lambda,\gamma) = \begin{cases}
        \lambda|u|&\mbox{if } |u|\le\lambda\\
        \frac{2\gamma\lambda|u|-u^2-\lambda^2}{2(\gamma-1)}&\mbox{if } \lambda\le |u| \le \gamma\lambda\\
        \frac{\lambda^2(\gamma+1)}{2}&\mbox{if } |u|>\gamma\lambda
    \end{cases},
\end{equation*}
for $\gamma>2$ and $\lambda>0$. The SCAD penalty is identical to the LASSO penalty for small coefficients, but continuously relaxes the rate of penalization as the coefficient departs from zero. Unlike OLS or LASSO, we have to solve a non-convex optimization problem that may have multiple minima and is computationaly more intensive than the LASSO. Nevertheless, \citet{jFlXhZ2014} showed how to calculate the oracle estimator using an iterative Local Linear Approximation algorithm.

\subsubsection*{Other Penalties}
Regularization imposes a restriction on the solution space, possibly imposing sparsity. In a data-rich environment it is a desirable property as it is likely that many regressors are not relevant to our prediction problem. The presentation above concentrates on the, possibly, most used penalties in time series forecasting. Nevertheless, there are many alternative penalties that can be used in regularized linear models.

% GROUP LASSO
The group LASSO, proposed by \citet{mYyL2006}, penalizes the parameters in groups, combining the $\ell_1$ and $\ell_2$ norms. It is motivated by the problem of identifying "factors", denoted by groups of regressors as, for instance, in regression with categorical variables that can assume many values. Let $\mathcal{G} = \{g_1,...,g_M\}$ denote a partition of $\{1,...,n\}$ and $\b\beta_{g_i} = [\beta_i:i\in g_i]$ the corresponding regression sub-vector. The group lasso assign to \eqref{E:shrinkage} the penalty  $p(\b\beta) = \sum_{i=1}^M\sqrt{|g_i|}\|\b\beta_{g_i}\|_2$, where $|g_i|$ is the cardinality of set $g_i$. The solution is efficiently estimated using, for instance, the group-wise majorization-descent algorithm \cite{yYhZ2015}. Naturally, the adaptive group LASSO was also proposed aiming to improve some of the limitations present on the group LASSO algorithm \cite{hWcL2008}. In the group LASSO, the groups enter or not in the regression. The sparse group LASSO recover sparse groups by combining the group LASSO penalty with the $L_1$ penalty on the parameter vector \citep{nSjFtHrT2013}.

\citet{hPfS2013} modify the adaptive lasso penalty to explicitly take into account lag information. \cite{eKfZ2016} propose a small change in penalty and perform a large simulation study to asses the performance of this penalty in distinct settings. They observe that taking into account lag information improves model selection and forecasting performance when compared to the LASSO and adaLASSO. They apply their method to forecasting inflation and risk premium with satisfactory results.

% BAYESIAN SHRINKAGE
There is a Bayesian interpretation to the regularization methods presented here. The ridge regression can be also seen as a maximum a posteriori estimator of a Gaussian linear regression with independent, equivariant, Gaussian priors. The LASSO replaces the Gaussian prior by a Laplace prior \citep{tPgC2008,cH2009}. These methods fall within the area of Bayesian Shrinkage methods, which is a very large and active research area, and it is beyond the scope of this survey.

\subsection{Theoretical properties}
% I Think we should be brief here. I would drop the definitions and stick with a simpler statement of results.
In this section we give an overview of the  theoretical properties of penalized regression estimators previously discussed. Most results in high-dimensional time series estimation focus on model selection consistency, oracle property and oracle bounds, for both the finite dimension ($n$ fixed, but possibly larger than $T$) and high-dimension ($n$ increases with $T$, usually faster).

More precisely, suppose there is a population, parameter vector $\b\beta_0$ that minimizes equation \eqref{eq:modely} over repeated samples. Suppose this parameter is sparse in a sense that only components indexed by $S_0\subset\{1,...,n\}$ are non-null. Let $\widehat{S_0}:=\{j:\widehat{\beta}_{j}\neq 0\}$. We say a method is \emph{model selection consistent} if the index of non-zero estimated components converges to $S_0$ in probability.\footnote{A more precise treatment would separate \emph{sign consistency} from \emph{model selection consistency}. \emph{Sign consistency} first appeared in \citet{pZbY2006} and also  verify whether the the sign of estimated regression weights converge to the population ones. }
\[\P(\widehat{S_0}=S_0)\to 1,\quad T\to \infty.\]
Consistency can also be stated in terms of how close the estimator is to true parameter for a given norm. We say that the estimation method is $\m{L}^q$-consistent if for every $\epsilon>0$:
\[\P(\|\widehat{\b\beta}_0-\b\beta_0\|_q>\epsilon)\to 0,\quad T\to \infty.\]
It is important to note that model selection consistency does not imply, nor it is implied by, $\m{L}^q$-consistency. As a matter of fact, one usually have to impose specific assumptions to achieve each of those modes of convergence.

Model selection performance of a given estimation procedure  can be further broke down in terms of how many relevant variables $j\in S_0$ are included in the model (screening). Or how many irrelevant variables $j\notin S_0$ are excluded from the model. In terms of probability, model screening consistency is defined by $\P(\widehat{S_0}\supseteq S_0)\to 1$ and model exclusion consistency defined by $\P(\widehat{S_0}\subseteq S_0)\to 1$ as $T\to\infty$.

We say a penalized estimator has the oracle property if its asymptotic distribution is the same as the unpenalized one only considering the $S_0$ regressors. Finally, oracle risk bounds are finite sample bounds on the estimation error of $\widehat{\b\beta}$ that hold with high probability. These bounds require relatively strong conditions on the curvature of objective function, which translates into a bound on the minimum  restricted eigenvalue of the covariance matrix among predictors for linear models and a rate condition on $\lambda$ that involves the number of non-zero parameters, $|S_0|$.

% LASSO
The LASSO was originally developed in fixed design with independent and identically distributed (IID) errors, but it has been extended and adapted to a large set of models and designs. \cite{kKwF2000} was probably the first paper to consider the asymptotics of the LASSO estimator. The authors consider fixed design and fixed $n$ framework. From their results, it is clear that the distribution of the parameters related to the irrelevant variables is non-Gaussian. To our knowledge, the first work expanding the results to a dependent setting was \cite{hWgLcT2007}, where the error term was allowed to follow an autoregressive process. Authors show that LASSO is model selection consistent, whereas a modified LASSO, similar to the adaLASSO, is both model selection consistent and has the oracle property. \cite{yNaR2011} shows model selection consistency and prediction consistency for lag selection in autoregressive models. \cite{kCkC2011} shows oracle properties and model selection consistency for lag selection in ARMA models. \cite{yYcPtL2013} derives model selection consistency and asymptotic distribution of the LASSO, adaLASSO and SCAD, for penalized regressions with autoregressive error terms. \citet{hSyS2015} studies lag estimation of autoregressive processes with long memory innovations using general penalties and show model selection consistency and asymptotic distribution for the LASSO and SCAD as particular cases. \citet{aK2016} shows model selection consistency and oracle property of adaLASSO for lag selection in stationary and integrated processes. All results above hold for the case of fixed number of regressors or relatively high-dimension, meaning that $n/T\to 0$.

In sparse, high-dimensional, stationary univariate time-series settings, where $n\to\infty$ at some rate faster than $T$,  \citet{mMeM2016, mMeM2017} show model selection consistency and oracle property of a large set of linear time series models with difference martingale, strong mixing, and non-Gaussian innovations. It includes,  predictive regressions, autoregressive models $AR(p)$, autoregressive models with exogenous variables $ARX(p)$, autoregressive models with dynamic lags $ADL(p,r)$, with possibly conditionally heteroscedastic errors. \citet{fXlXyY2017} shows oracle bounds for fixed design regression with $\beta$-mixing errors. \citet{wWyW2016} derive oracle bounds for the LASSO on regression with fixed design and weak dependent innovations, in a sense of \citet{wW2005}, whereas \citet{yHrT2020} show model selection consistency for linear regression with random design and weak sparsity\footnote{\emph{Weak sparsity} generalizes sparsity by supposing that coefficients are (very) small instead of exactly zero.} under serially dependent errors and covariates, within the same weak dependence framework. \cite{yXmT2020} show model selection consistency and parameter consistency for a modified version of the LASSO in time series regressions with long memory innovations.

\citet{jFrL2001} shows model selection consistency and oracle property for the folded concave penalty estimators in a fixed dimensional setting. \citet{yKhChO2008} showed that the SCAD also enjoys these properties in high-dimensions. In time-series settings,\citet{yUsT2019} shows oracle properties and model selection consistency in time series models with dependent regressors. \cite{jLlYiG2019} derived oracle prediction bounds for many penalized regression problems. The authors conclude that generic high dimensional penalized estimators provide consistent prediction with any design matrix. Although the results are not directly focused on time series problems, they are general enough to hold in such setting.

\citet{aBeGjS2020a} proposed the sparse-group LASSO as an estimation technique when high-dimensional time series data are potentially sampled at different frequencies. The authors derived oracle inequalities for the sparse-group LASSO estimator within a framework where distribution of the data may have heavy tails.

Two frameworks not directly considered in this survey but of great empirical relevance are nonstationary environments and multivariate models. In sparse, high-dimensional, integrated time series settings, \citet{jLzSzG2018} and \citet{bKhAmSwY2020} show model selection consistency and derive the asymptotic distributions of LASSO estimators and some variants. \citet{sSeW2020} proposed the Single-equation Penalized Error Correction Selector (SPECS), which is an automated estimation procedure for dynamic single-equation models with a large number of potentially co-integrated variables. In sparse multivariate time series, \cite{nHnNyC2008} shows model selection consistency in VAR models with white-noise shocks. \citet{yRxZ2010} uses adaLASSO in a similar setting, showing both model selection consistency and oracle property. Afterwards, \citet{lCaK2014} show model selection consistency and oracle property of the adaptive Group LASSO. In high dimensional settings, where the dimension of the series increase with the number of observations, \citet{aKlC2015,sBgM2015} shows oracle bounds and model selection consistency for the LASSO in Gaussian $VAR(p)$ models, extending previous works. \citet{iMaB2016} extended these results for a large collection of penalties. \citet{xZ2020} derive oracle estimation bounds for folded concave penalties for Gaussian $VAR(p)$ models in high dimensions. More recently researchers have departed from gaussianity and correct model specification. \citet{kWzLaT2020} derived finite-sample guarantees for the LASSO in a misspecified VAR model involving $\beta$-mixing process with sub-Weibull marginal distributions. \citet{rMmMeM2019} derive equation-wise error bounds for the LASSO estimator of weakly sparse $VAR(p)$ in mixingale dependence settings, that include models with conditionally heteroscedastic innovations.

\subsection{Inference}
Although several papers derived the asymptotic properties of penalized estimators as well as the oracle property, these results have been derived under the assumption that the true non-zero coefficients are large enough. This condition is known as the $\b\beta$-min restriction. Furthermore, model selection, such as the choice of the penalty parameter, has not been taken into account. Therefore, the true limit distribution, derived under uniform  asymptotics and without the $\b\beta$-min restriction can bee very different from Gaussian, being even bimodal; see, for instance, \cite{leeb2005model}, \cite{leeb2008sparse}, and \cite{aBvCcH2014a} for a detailed discussion.

Inference after model selection is actually a very active area of research and a vast number of papers have recently appeared in the literature. \cite{sGpByRrD2014} proposed the desparsified LASSO in order to construct (asymptotically) a valid confidence interval for each $\beta_{j,0}$ by modifying the original LASSO estimate $\widehat{\b\beta}$. Let $\b\Sigma^*$ be an approximation for the inverse of $\b\Sigma:=\E(\b X_t \b X_t')$,  then the desparsified LASSO is defined as $\widetilde{\b\beta}:=\widehat{\b\beta}+ \b\Sigma^*(\b Y- \b X\widehat{\b\beta})/T$.  The addition of this extra term to the LASSO estimator results in an unbiased estimator that no longer estimate any coefficient exactly as zero. More importantly, asymptotic normality can be recover in the sense that $\sqrt{T}(\widetilde{\beta}_i - \beta_{i,0})$ converges in distribution to a Gaussian distribution under appropriate regularity conditions. Not surprisingly, the most important condition is how well $\b\Sigma^{-1}$ can be approximated by $\b\Sigma^*$. In particular, the authors propose to run $n$ LASSO regressions of $X_i$ onto $\b X_{-i}:=(X_1,\dots, X_{i-1},X_{i+1},\dots, X_n)$, for $1\leq i\leq n$. The authors named this process as \emph{nodewide regressions}, and use those estimates to construct $\b\Sigma^*$ (refer to Section 2.1.1 in \cite{sGpByRrD2014} for details).

\citet{aBvCcH2014a} put forward the double-selection method in the context of on a linear model in the form $Y_t= \beta_{01} X_{t}^{(1)} + \b\beta'_{02}\b X^{(2)}_{t} + U_t$, where the interest lies on the the scalar parameter $\beta_{01}$ and $\b X_t^{(2)}$ is a high-dimensional vector of control variables. The procedure consists in obtaining an estimation of the active (relevant) regressors in the high-dimension auxiliary regressions of $Y_t$ on $\b X^{(2)}$ and of $X_t^{(1)}$ on  $\b X_t^{(2)}$, given by $\widehat{S}_1$ and $\widehat{S}_2$, respectively.\footnote{The relevant regressors are the ones associated with non-zero parameter estimates.} This can be obtained either by LASSO or any other estimation procedure. Once the set $\widehat{S}:=\widehat{S}_1\cup\widehat{S}_2$ is identified, the (a priori) estimated non-zero parameters can by estimated by a low-dimensional regression $Y_t$ on $X_t^{(1)}$ and $\{X_{it}^{(2)}:i\in\widehat{S}\}$. The main result (Theorem 1 of \citet{aBvCcH2014a}) states conditions under which the estimator $\widehat{\beta}_{01}$ of the parameter of interest properly studentized is asymptotically normal. Therefore, uniformly valid asymptotic confidence intervals for $\beta_{01}$ can be constructed in the usual fashion.

Similar to \citet{jTrLrjTrT2014} and \citet{rLjTrTtT2014}, \citet{jdLdlSySjeT2016} put forward general approach to valid inference after model selection. The idea is to characterize the distribution of a post-selection estimator conditioned on the selection event. More specifically, the authors argue that the post-selection confidence intervals for regression coefficients should have the correct coverage conditional on the selected model. The specific case of the LASSO estimator is discussed in details. The main difference between \citet{jdLdlSySjeT2016} and \citet{jTrLrjTrT2014} and \citet{rLjTrTtT2014} is that in the former, confidence intervals can be formed at any value of the LASSO penalty parameter and any coefficient in the model. Finally, it is important to stress that \citet{jdLdlSySjeT2016} inference is carried on the coefficients of the selected model, while \cite{sGpByRrD2014} and \citet{aBvCcH2014a} consider inference on the coefficients of the true model.

The above papers do not consider a time-series environment. \citet{aHlMsS2019} is on the first papers which attempt to consider post-selection inference in a time-series environment. The authors generalize the results in \citet{aBvCcH2014a} to dependent processes. However, their results are derived under a fixed number of variables. \citet{aBeGjS2020c} and \cite{rAsSiW2020} extend the seminal work of \citet{sGpByRrD2014} to time-series framework. More specifically, \citet{aBeGjS2020c} consider inference in time-series regression models under heteroskedastic and autocorrelated errors. The authors consider heteroskedaticity- and autocorrelation-consistent (HAC) estimation with sparse group-LASSO. They propose a debiased central limit theorem for low dimensional groups of regression coefficients and study the HAC estimator of the long-run variance based on the sparse-group LASSO residuals. \cite{rAsSiW2020} extend the desparsified LASSO to a time-series setting under near-epoch dependence assumptions, allowing for non-Gaussian, serially correlated and heteroskedastic processes. Furthermore, the number of regressors can possibly grow faster than the sample size.

\section{Nonlinear Models}\label{S:nonlinear}

The function $f_h$ appearing \eqref{eq:model} is unknown and in several applications the linearity assumption is too restrictive and more flexible forms must be considered. Assuming a quadratic loss function, the estimation problem turns to be the minimization of the functional
\begin{equation}\label{E:inf}
S(f):=\sum_{t=1}^{T-h}\left[Y_{t+h}-f(\b{X}_t)\right]^2,
\end{equation}
where $f\in \mathcal{G}$, a generic function space. However, the optimization problem stated in \eqref{E:inf} is infeasible when $\mathcal{G}$ is infinite dimensional, as there is no efficient technique to search over all $\mathcal{G}$. Of course, one solution is to restrict the function space, as for instance, imposing linearity or specific forms of parametric nonlinear models as in, for example, \citet{tT1994a}, \citet{sMcPmcM2004} or \citet{mMmcM2008}; see also \cite{tTdTcwjG2010} for a recent review of such models.

Alternatively, we can replace $\mathcal{G}$ by simpler and finite dimensional $\mathcal{G}_D$. The idea is to consider a sequence of finite dimensional spaces, the \emph{sieve} spaces,  $\mathcal{G}_{D},\,D=1,2,3,\ldots,$ that converges to $\mathcal{G}$ in some norm. The approximating function $g_D(\b{X}_t)$ is written as
\[
g_D(\b{X}_t)=\sum_{j=1}^{J}\beta_jg_j(\b{X}_t),
\]
where $g_j(\cdot)$ is the $j$-th \emph{basis} function for $\mathcal{G}_{D}$ and can be either fully known or indexed by a vector of parameters, such that: $g_j(\b{X}_t):=g(\b{X}_t;\b{\theta}_j)$. The number of basis functions $J:=J_T$ will depend on the sample size $T$. $D$ is the dimension of the space and it also depends on the sample size: $D:=D_T$. Therefore, the optimization problem is then modified to
\begin{equation}\label{E:feas}
\widehat{g}_D(\b{X}_t)=\arg\underset{g_D(\b{X}_t)\in\mathcal{G}_D}{\min}\sum_{t=1}^{T-h}\left[Y_{t+h}-g_D(\b{X}_t)\right]^2.
\end{equation}

The sequence of approximating spaces $\mathcal{G}_{D}$ is chosen by using the structure of the original underlying space $\mathcal{G}$ and the fundamental concept of dense sets. If we have two sets $A$ and $B$ $\in \mathcal{X}$, $\mathcal{X}$ being a metric space, $A$ is dense in $B$ if for any $\epsilon>0, \in \R$ and $x \in B$ there is a $y \in A$ such that $\|x-y\|_{\mathcal{X}}< \epsilon$. This is called the method of \textbf{sieves}. For a comprehensive review of the method for time-series data, see \citet{cX2007}.

For example, from the theory of approximating functions we know that the proper subset $\mathcal{P} \subset \mathcal{C}$ of polynomials is dense in $\mathcal{C}$, the space of continuous functions. The set of polynomials is smaller and simpler than the set of all continuous functions. In this case, it is natural to define the sequence of approximating spaces $\mathcal{G}_{D},\,D=1,2,3,\ldots$ by making $\mathcal{G}_{D}$ the set of polynomials of degree smaller or equal to $D-1$ (including a constant in the parameter space). Note that $\mathsf{dim}(\mathcal{G}_{D})= D < \infty$. In the limit this sequence of finite dimensional spaces converges to the infinite dimensional space of polynomials, which on its turn is dense in $\mathcal{C}$.

When the basis functions are all known (\textbf{linear sieves}), the problem is linear in the parameters and methods like ordinary least squares (when $J\ll T$) or penalized estimation as previously described can be used.

For example, let $p=1$ and pick a polynomial basis such that
\[
g_D(X_t)=\beta_0 +\beta_1X_t +\beta_2X_t^2+\beta_3X_t^3+\cdots+\beta_JX_t^{J}.
\]
In this case, the dimension $D$ of $\mathcal{G}_D$ is $J+1$, due to the presence of a constant term.

If $J<<T$, the vector of parameters $\b\beta=(\beta_1,\ldots,\beta_J)'$ can be estimated by
\[
\widehat{\b\beta}=\left(\b{X}_J'\b{X}_J\right)^{-1}\b{X}_J'\b{Y},
\]
where $\b{X}_J$ is the $T\times (J+1)$ design matrix and $\b{Y}=(Y_1,\ldots,Y_T)'$.

When the basis functions are also indexed by parameters (\textbf{nonlinear sieves}), nonlinear least-squares methods should be used. In this paper we will focus on frequently used nonlinear sieves: neural networks and regression trees.

\subsection{Neural Networks}\label{S:NN}
\subsubsection{Shallow Neural Networks}
Neural Networks (NN) is one of the most traditional nonlinear sieves. NN can be classified into shallow or deep networks. We start describing the shallow NNs. The most common shallow NN is the feedforward neural network where the the approximating function $g_D(\b{X}_t)$ is defined as
\begin{equation}\label{E:NN}
\begin{split}
g_D(\b{X}_t):=g_D(\b{X}_t;\b\theta)&=\beta_0+\sum_{j=1}^{J_T}\beta_jS(\b{\gamma}_j'\b{X}_t+\gamma_{0,j}),\\
&=\beta_0+\sum_{j=1}^{J_T}\beta_j S(\tilde{\b{\gamma}}_j'\tilde{\b{X}}_t),
\end{split}
\end{equation}
In the above model, $\tilde{\b{X}}_t=(1,\b{X}_t')'$, $S_j(\cdot)$ is a basis function and the parameter vector to be estimated is given by  $\b\theta=(\beta_0,\ldots,\beta_K,\b\gamma_1',\ldots,\b\gamma_{J_T}',\gamma_{0,1},\ldots,\gamma_{0,J_T})'$, where  $\tilde{\b{\gamma}}_j=(\gamma_{0,j},\b\gamma_j')'$.

\begin{figure}
\centering
\includegraphics[width=0.6\linewidth]{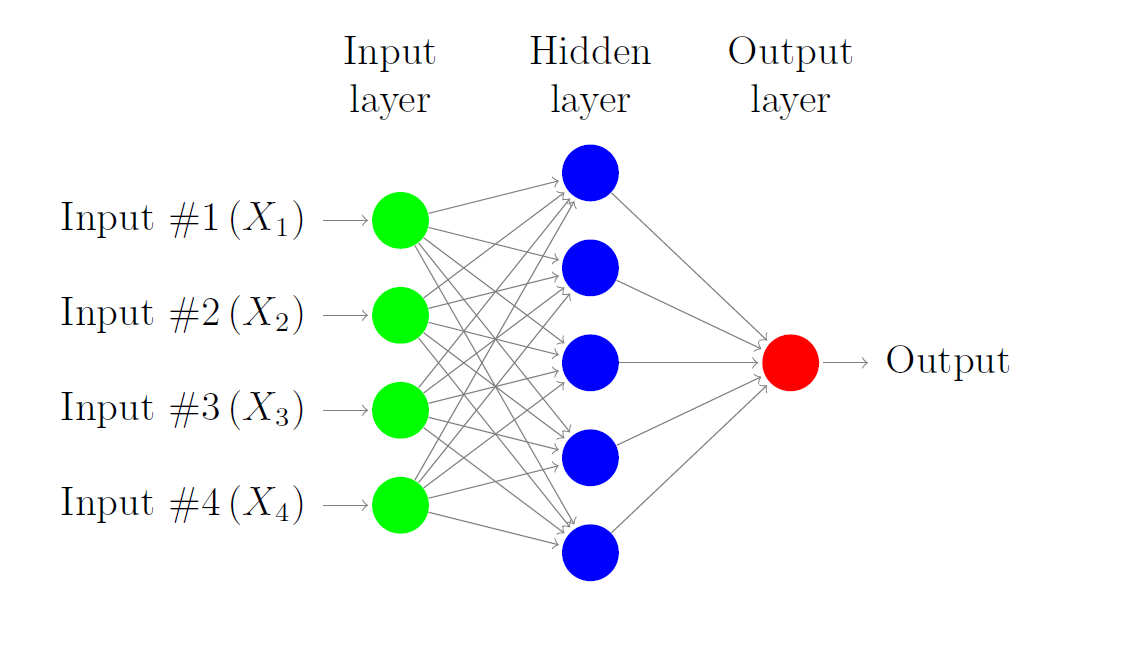}
\caption{Graphical representation of a single hidden layer neural network.}
\label{F:NN}
\end{figure}

NN models form a very popular class of nonlinear sieves and have been used in many applications of economic forecasting. Usually, the basis functions $S(\cdot)$ are called activation functions and the parameters are called weights. The terms in the sum are called hidden-neurons as an unfortunate analogy to the human brain. Specification \eqref{E:NN} is also known as a single hidden layer NN model as is usually represented in the graphical as in Figure \ref{F:NN}. The green circles in the figure represent the input layer which consists of the covariates of the model ($\b{X}_t$). In the example in the figure there are four input variables. The blue and red circles indicate the hidden and output layers, respectively. In the example, there are five elements (neurons) in the hidden layer.The arrows from the green to the blue circles represent the linear combination of inputs: $\b{\gamma}_j'\b{X}_t+\gamma_{0,j}$, $j=1,\ldots,5$. Finally, the arrows from the blue to the red circles represent the linear combination of outputs from the hidden layer: $\beta_0+\sum_{j=1}^{5}\beta_jS(\b{\gamma}_j'\b{X}_t+\gamma_{0,j})$.

There are several possible choices for the activation functions. In the early days, $S(\cdot)$ was chosen among the class of squashing functions as per the definition bellow.

\begin{Def}
A function $S:\mathbb{R}\longrightarrow[a,b]$, $a<b$, is a squashing (sigmoid) function if it is non-decreasing, $\underset{x\longrightarrow\infty}{\lim}S(x)=b$ and $\underset{x\longrightarrow-\infty}{\lim}S(x)=a$.
\end{Def}

Historically, the most popular choices are the logistic and hyperbolic tangent functions such that:
\[
\begin{split}
\textnormal{Logistic: }&S(x)=\frac{1}{1+\exp(-x)}\\
\textnormal{Hyperbolic tangent: }&S(x)=\frac{\exp(x)-\exp(-x)}{\exp(x)+\exp(-x)}.
\end{split}
\]

The popularity of such functions was partially due to theoretical results on function approximation. \citet{kF1989} establishes that NN models as in \eqref{E:NN} with generic squashing functions are capable of approximating any continuous functions from one finite dimensional space to another to any desired degree of accuracy, provided that $J_T$ is sufficiently large. \citet{gC1989} and \citet{kHmShW1989} simultaneously proved approximation capabilities of NN models to any Borel measurable function and \citet{kHmShW1989} extended the previous results and showed the NN models are also capable to approximate the derivatives of the unknown function. \citet{aB1993} relate previous results to the number of terms in the model.

\citet{mSwH1989} and \citet{jPiwS1991} derived the same results of \citet{gC1989} and \citet{kHmShW1989} but without requiring the activation function to be sigmoid. While the former considered a very general class of functions, the later focused on radial-basis functions (RBF) defined as:
\[
\textnormal{Radial Basis: } S(x)=\exp(-x^2).
\]

More recently, \citet{dY2017} showed that the rectified linear units (ReLU) as
\[
\textnormal{Rectified Linear Unit: }S(x)=\max(0,x),
\]
are also universal approximators.

Model \eqref{E:NN} can be written in matrix notation. Let $\b\Gamma=(\tilde{\b\gamma}_1,\ldots,\tilde{\b\gamma}_K)$,
\[
\b{X}=
\begin{pmatrix}
1 & X_{11} & \cdots & X_{1p}\\
1 & X_{21} & \cdots & X_{2p}\\
\vdots &  \ddots & \vdots \\
1 & X_{T1} & \cdots & X_{Tp}\\
\end{pmatrix},\,\textnormal{and}\,
\b{\mathcal{O}}(\b{X}\b{\Gamma})=
\begin{pmatrix}
1&S(\tilde{\b\gamma}_1'\tilde{\b{x}}_1) & \cdots & S(\tilde{\b\gamma}_K'\tilde{\b{x}}_1)\\
1&S(\tilde{\b\gamma}_1'\tilde{\b{x}}_2) & \cdots & S(\tilde{\b\gamma}_K'\tilde{\b{x}}_2)\\
\vdots &\vdots & \ddots & \vdots \\
1&S(\tilde{\b\gamma}_1'\tilde{\b{x}}_T) & \cdots & S(\tilde{\b\gamma}_K'\tilde{\b{x}}_T)\\
\end{pmatrix}
\]
Therefore, by defining $\b\beta=(\beta_0,\beta_1,\ldots,\beta_K)'$, the output of a feed-forward NN is given by:
\begin{equation}\label{E:NNmat}
\begin{split}
\b{h}_D(\b{X},\b\theta)&=[h_D(\b{X}_1;\b\theta),\ldots,h_D(\b{X}_T;\b\theta)]'\\
&=
\begin{bmatrix}
\beta_0+\sum_{k=1}^K\beta_k S(\b\gamma_k'\b{X}_1+\gamma_{0,k})\\
\vdots\\
\beta_0+\sum_{k=1}^K\beta_k S(\b\gamma_k'\b{X}_T+\gamma_{0,k})\\
\end{bmatrix}\\
&=\b{\mathcal{O}}(\b{X}\b{\Gamma})\b\beta.
\end{split}
\end{equation}

The dimension of the parameter vector $\b{\theta}=[\vec(\b\Gamma)',\b\beta']'$ is $k=(n+1)\times J_T +(J_T+1)$ and can easily get very large such that the unrestricted estimation problem defined as
\[
\widehat{\b\theta}=\arg\underset{\b\theta\in\R^k}{\min}\|\b{Y}-\mathcal{O}(\b{X}\b\Gamma)\b\beta\|_2^2
\]
is unfeasible. A solution is to use regularization as in the case of linear models and consider the minimization of the following function:
\begin{equation}\label{E:costNN}
Q(\b\theta)=\|\b{Y}-\mathcal{O}(\b{X}\b\Gamma)\b\beta\|_2^2 +p(\b\theta),
\end{equation}
where usually $p(\b\theta)=\lambda\b\theta'\b\theta$. Traditionally, the most common approach to minimze \eqref{E:costNN} is to use Bayesian methods as in \citet{djcM1992}, \cite{djcM1992b}, and \citet{fdFmtH1997}. A more modern approach is to use a technique known as \emph{Dropout} \citep{nS2014}.

The key idea is to randomly drop neurons (along with their connections) from the neural network during estimation. A NN with $J_T$ neurons in the hidden layer can generate $2^{J_T}$ possible ``thinned'' NN by just removing some neurons. Dropout samples from this  $2^{J_T}$ different thinned NN and train the sampled NN. To predict the target variable, we use a single unthinned network that has weights adjusted by the probability law induced by the random drop. This procedure significantly reduces overfitting and gives major improvements over other regularization methods.

We modify equation \eqref{E:NN} by
\[
g_D^*(\b{X}_t)=\beta_0+\sum_{j=1}^{J_T}s_j\beta_jS(\b{\gamma}_j'\left[\b{r}\odot\b{X}_t\right]+v_j\gamma_{0,j}),
\]
where $s$, $v$, and $\b{r}=(r_1,\ldots,r_n)$ are  independent Bernoulli random variables each with probability $q$ of being equal to $1$. The NN model is thus estimated by using  $g_D^*(\b{X}_t)$ instead of $g_D(\b{X}_t)$ where, for each training example, the values of the entries of $\b{r}$ are drawn from the Bernoulli distribution. The final estimates for $\beta_j$, $\b\gamma_j$, and $\gamma_{o,j}$ are multiplied by $q$.

\subsubsection{Deep Neural Networks}
\begin{figure}
\centering
\includegraphics[width=0.6\linewidth]{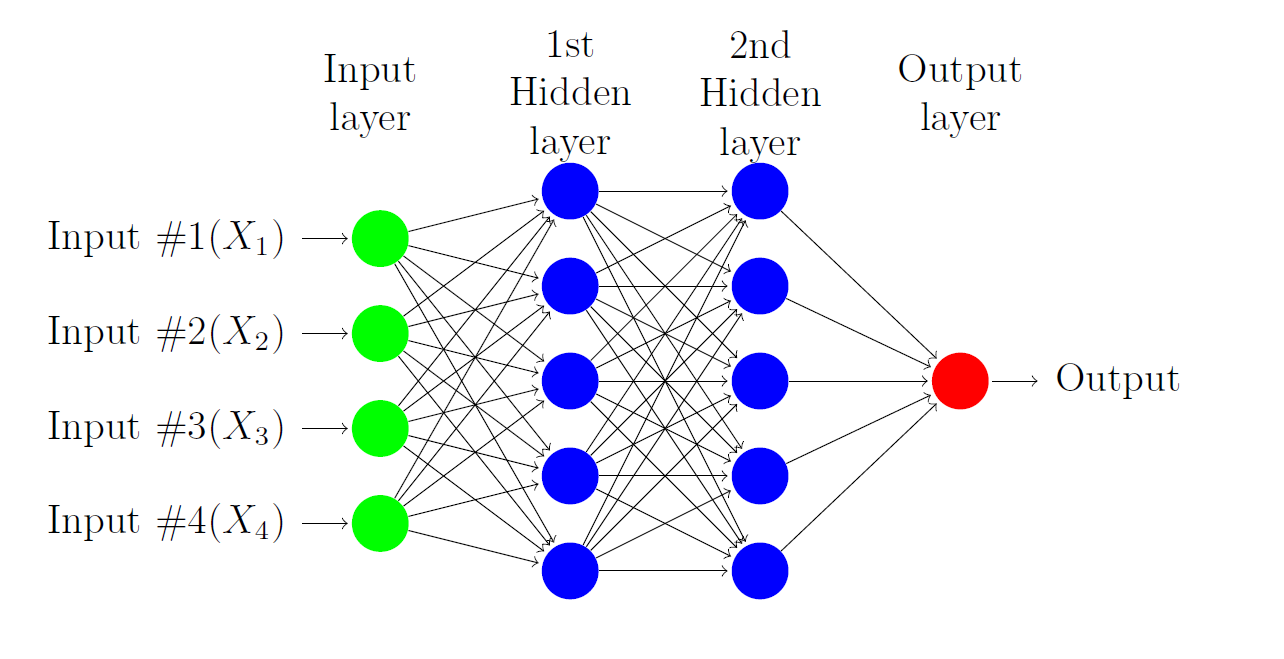}
\caption{Deep neural network architecture}
\label{F:DNN}
\end{figure}

A Deep Neural Network model is a straightforward generalization of specification \eqref{E:NN} where more hidden layers are included in the model as represented in Figure \ref{F:DNN}. In the figure we represent a Deep NN with two hidden layers with the same number of hidden units in each. However, the number of hidden neurons can vary across layers.

As pointed out in \citet{hMqLtP2017}, while the universal approximation property holds for shallow NNs, deep networks can approximate the class of compositional functions as well as shallow networks but with exponentially lower number of training parameters and sample complexity.

Set $J_{\ell}$ as the number of hidden units in layer $\ell\in\{1,\dots,L\}$. For each hidden layer $\ell$ define $\b\Gamma_{\ell}=(\tilde{\b\gamma}_{1\ell},\ldots,\tilde{\b\gamma}_{k_{\ell}\ell})$. Then the output $\mathcal{O}_\ell$ of layer $\ell$ is given recursively by
\[
\underset{n\times (J_{\ell}+1)}{\b{\mathcal{O}}_{\ell}(\b{\mathcal{O}}_{\ell-1}(\cdot)\b{\Gamma}_{\ell})}
=
\begin{pmatrix}
1&S(\tilde{\b\gamma}_{1\ell}'\b{\mathcal{O}}_{1\ell-1}(\cdot)) & \cdots & S(\tilde{\b\gamma}_{k_{\ell}\ell}'\b{\mathcal{O}}_{1\ell-1}(\cdot))\\
1&S(\tilde{\b\gamma}_{1\ell}'\b{\mathcal{O}}_{2\ell-1}(\cdot)) & \cdots & S(\tilde{\b\gamma}_{k_{\ell}\ell}'\b{\mathcal{O}}_{2\ell-1}(\cdot))\\
\vdots &\vdots & \ddots & \vdots \\
1&S(\tilde{\b\gamma}_{1\ell}'\b{\mathcal{O}}_{n\ell-1}(\cdot)) & \cdots & S(\tilde{\b\gamma}_{J_{\ell}\ell}'\b{\mathcal{O}}_{n\ell-1}(\cdot))\\
\end{pmatrix}
\]
where $\b{\mathcal{O}}_o:=\b{X}$. Therefore, the output of the Deep NN is the composition
\[
\b{h}_D(\b{X})=\b{\mathcal{O}}_{L}(
\cdots
\b{\mathcal{O}}_3(\b{\mathcal{O}}_2(\b{\mathcal{O}}_{1}(\b{X}\b\Gamma_1)\b\Gamma_2)\b\Gamma_3)
\cdots)\b\Gamma_L\b{\beta}.
\]

The estimation of the parameters is usually carried out by stochastic gradient descend methods with dropout to control the complexity of the model.

\subsubsection{Recurrent Neural Networks}
Broadly speaking, Recurrent Neural Networks (RNNs) are NNs that allow for feedback among the hidden layers. RNNs can use their internal state (memory) to process sequences of inputs. In the framework considered in this paper, a generic RNN could be written as
\[
\begin{split}
\b{H}_t &= \b f(\b{H}_{t-1},\b{X}_t),\\
\widehat{Y}_{t+h|t} &= g(\b{H}_t),
\end{split}
\]
where $\widehat{Y}_{t+h|t}$ is the prediction of $Y_{t+h}$ given observations only up to time $t$,  $\b f$ and $g$ are functions to be defined and $\b H_t$ is what we call the (hidden) state. From a time-series perspective, RNNs can be see as a kind of nonlinear state-space model.

\begin{figure}
    \centering
    \includegraphics[width=0.4\linewidth]{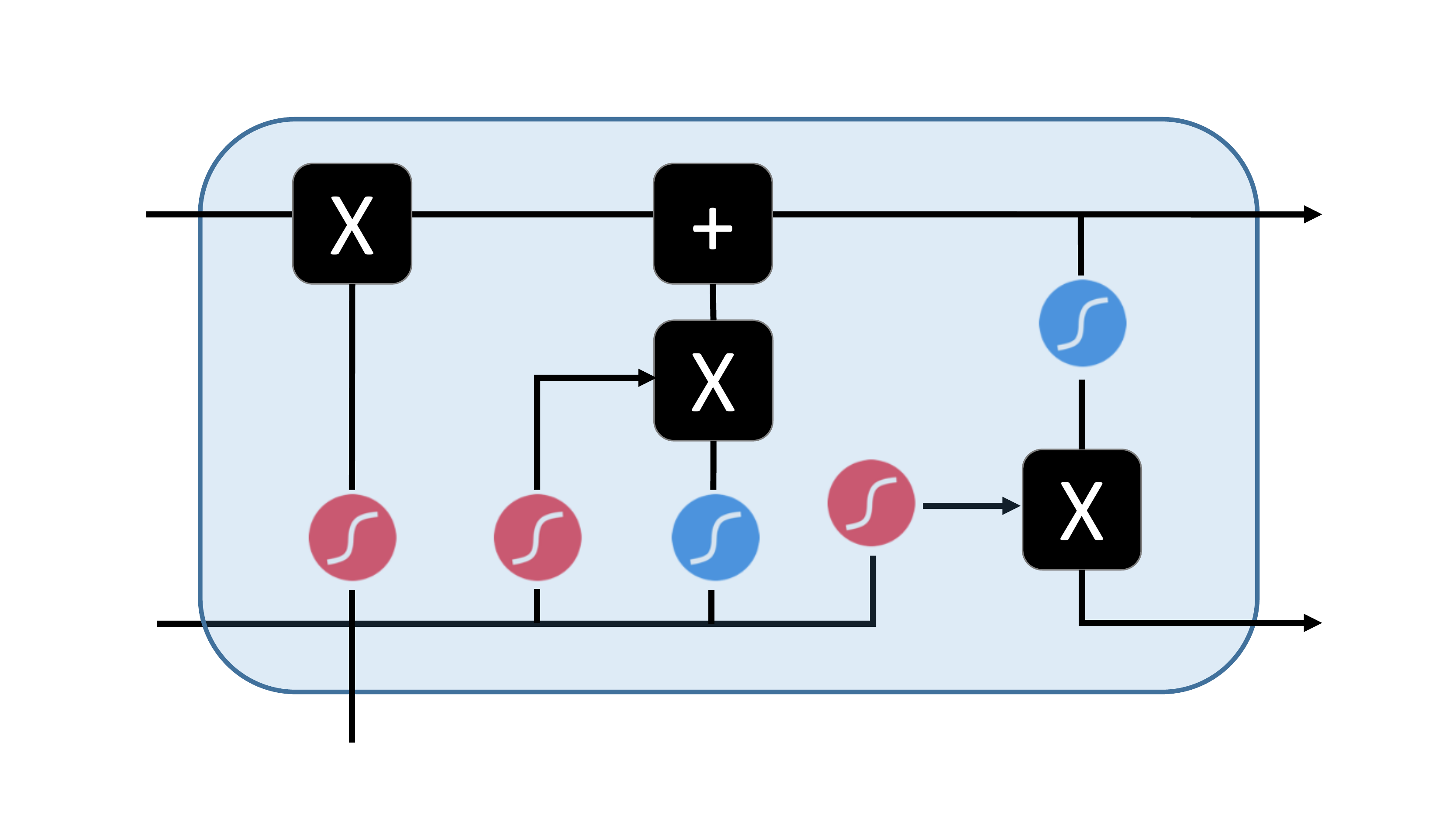}
    \includegraphics[width=0.4\linewidth]{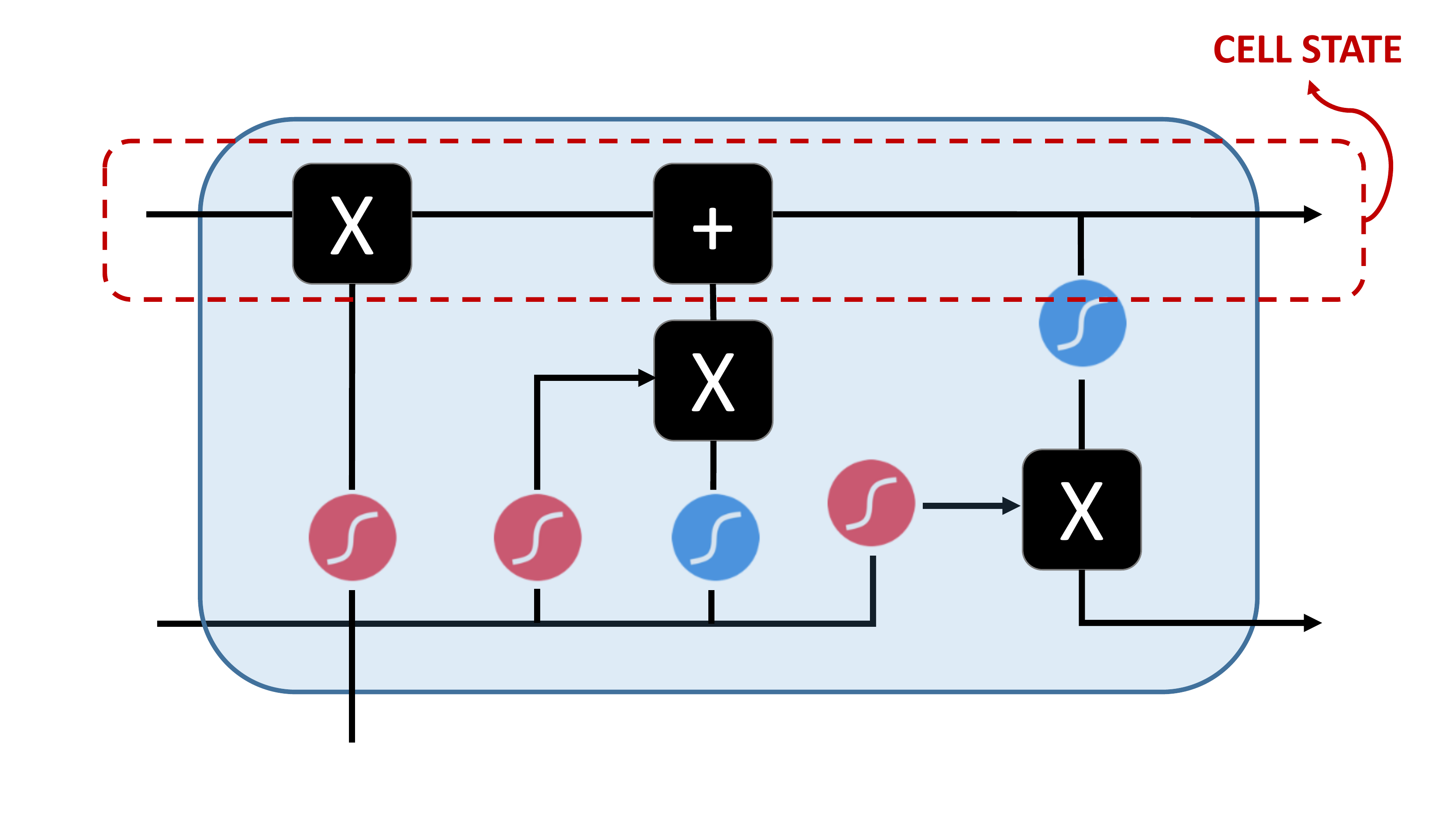}\\
    \includegraphics[width=0.4\linewidth]{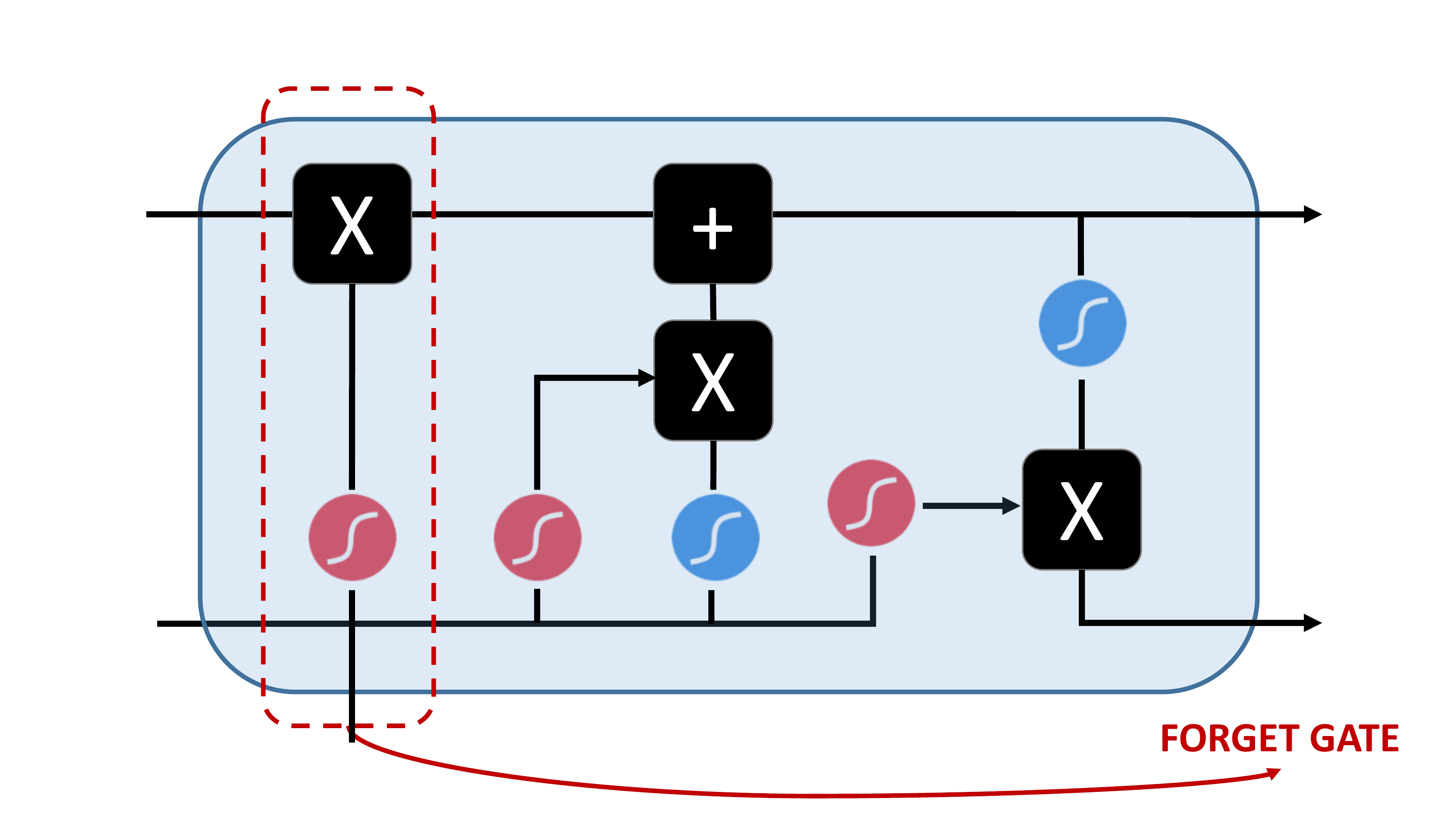}
    \includegraphics[width=0.4\linewidth]{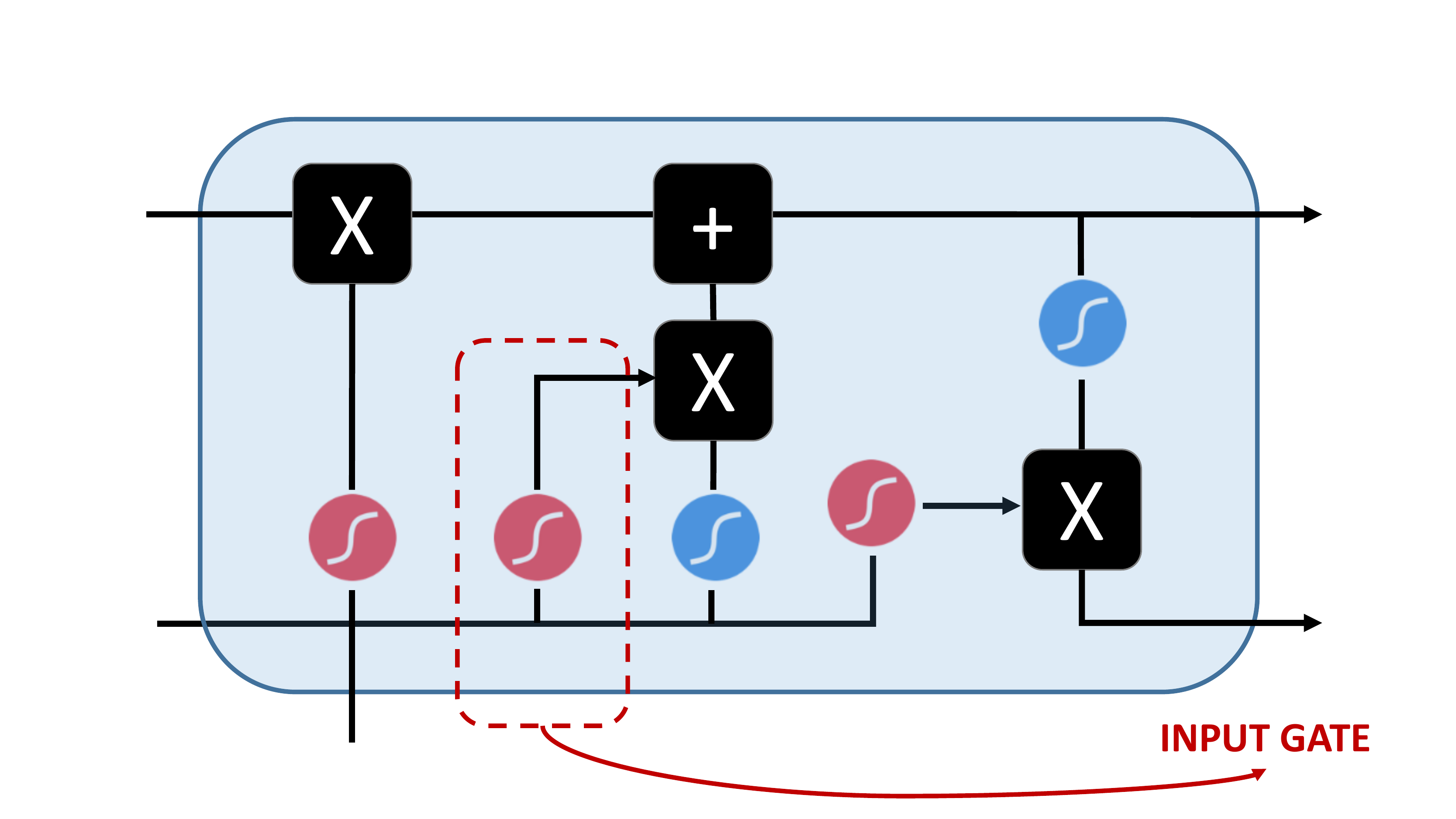}\\
    \includegraphics[width=0.4\linewidth]{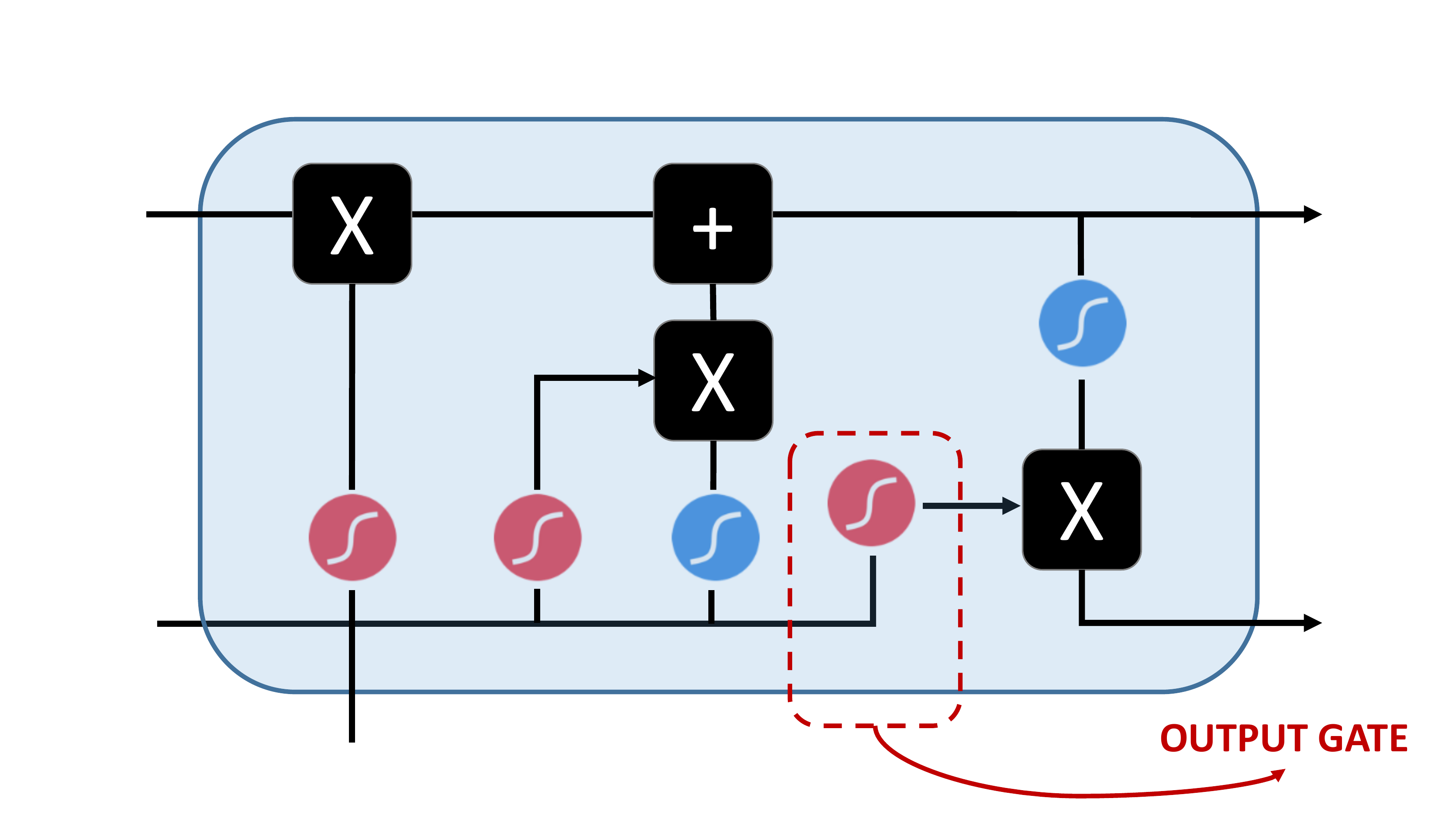}
    \caption{Architecture of the Long-Short-Term Memory Cell (LSTM)}
    \label{F:LSTM}
\end{figure}

RNNs can remember the order that the inputs appear through its hidden state (memory) and they can also model sequences of data so that each sample can be assumed to be dependent on previous ones, as in time series models. However, RNNs are hard to be estimated as they suffer from the vanishing/exploding gradient problem. Set the cost function to be
\[
\mathcal{Q}_T(\b\theta)=\sum_{t=1}^{T-h}\left(Y_{t+h}-\widehat{Y}_{t+h|t}\right)^2,
\]
where $\b\theta$ is the vector of parameters to be estimated. It is easy to show that the gradient $\frac{\partial \mathcal{Q}_T(\b\theta)}{\partial \b\theta}$ can be very small or diverge. Fortunately, there is a solution to the problem proposed by \citet{sHjS1997}. A variant of RNN which is called Long-Short-Term Memory (LSTM) network . Figure \ref{F:LSTM} shows the architecture of a typical LSTM layer. A LSTM network can be composed of several layers. In the figure, red circles indicate logistic activation functions, while blue circles represent hyperbolic tangent activation. The symbols ``\textsf{X}'' and ``\textsf{+}'' represent, respectively, the element-wise multiplication and sum operations. The RNN layer is composed of several blocks: the cell state and the forget, input, and ouput gates. The cell state introduces a bit of memory to the LSTM so it can ``remember'' the past. LSTM learns to keep only relevant information to make predictions, and forget non relevant data. The forget gate tells which information to throw away from the cell state. The output gate provides the activation to the final output of the LSTM block at time $t$. Usually, the dimension of the hidden state ($\b{H}_t$) is associated with the number of hidden neurons.

\begin{figure}
    \centering
    \includegraphics[width=0.9\linewidth]{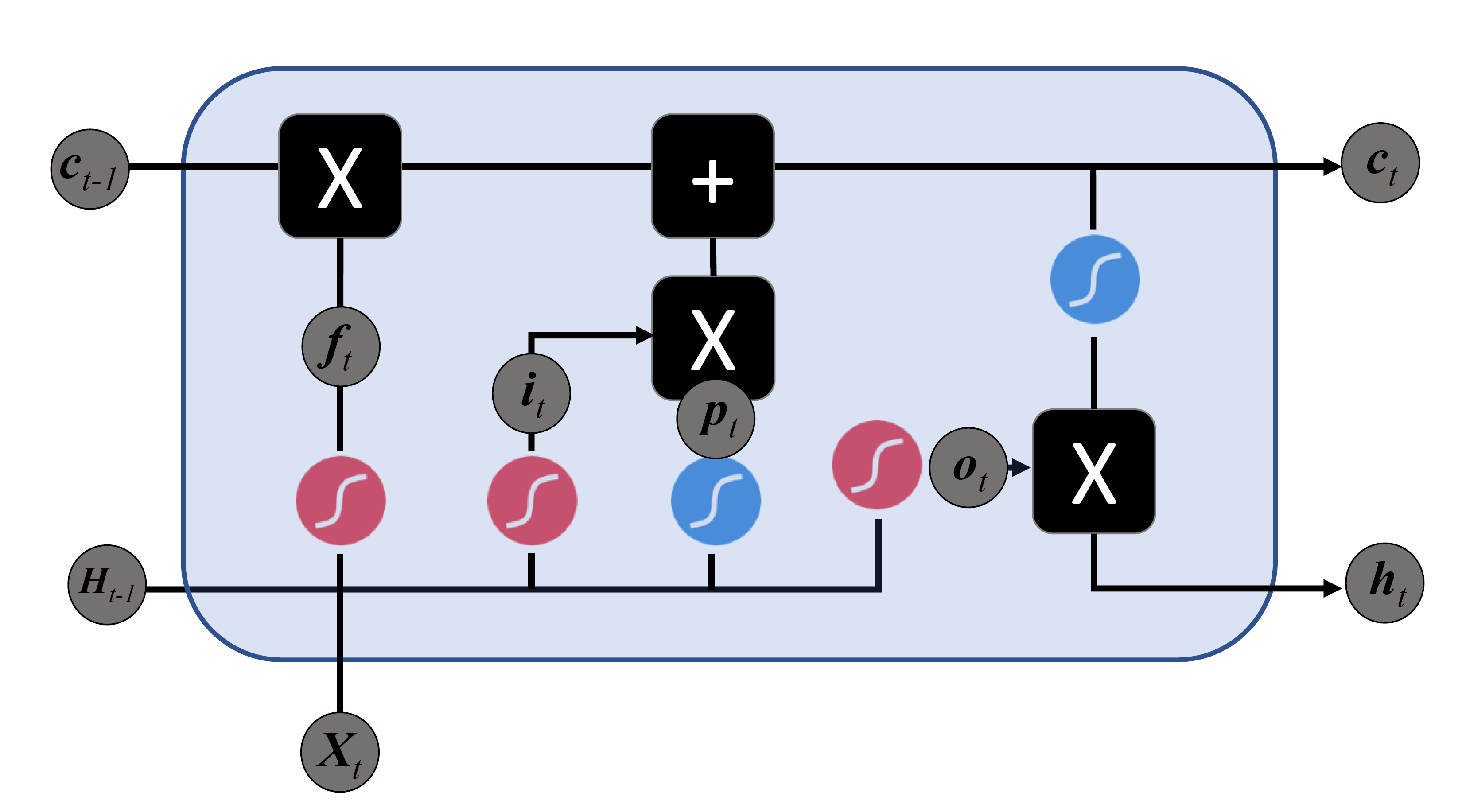}
    \caption{Information flow in a LTSM Cell}
    \label{F:LSTM2}
\end{figure}

Algorithm \ref{A:LSTM} describes analytically how the LSTM cell works. $\b f_t$ represents the output of the forget gate. Note that it is a combination of the previous hidden-state ($\b H_{t-1}$) with the new information ($\b X_t$). Note that $\b f_t\in[0,1]$ and it will attenuate the signal coming com $\b c_{t-1}$. The input and output gates have the same structure. Their function is to filter the ``relevant'' information from the previous time period as well as from the new input.  $\b p_{t}$ scales the combination of inputs and previous information. This signal will be then combined with the output of the input gate ($\b i_t$). The new hidden state will be an attenuation of the signal coming from the output gate. Finally, the prediction is a linear combination of hidden states. Figure \ref{F:LSTM2} illustrates how the information flows in a LSTM cell.

\begin{algorithm}\label{A:LSTM}
Mathematically, RNNs can be defined by the following algorithm:
\begin{enumerate}
\item
Initiate with $\b{c}_0=0$ and $\b{H}_0=0$.
\item
Given the input $\b{X}_t$, for $t\in\{1,\dots,T\}$, do:
\[
\begin{split}
\b{f}_t &= \textnormal{Logistic}(\b{W}_f \b{X}_t + \b{U}_f \b{H}_{t-1}+ \b{b}_f)\\
\b{i}_t &= \textnormal{Logistic}(\b{W}_i \b{X}_t + \b{U}_i \b{H}_{t-1}+ \b{b}_i)\\
\b{o}_t &= \textnormal{Logistic}(\b{W}_o \b{X}_t + \b{U}_o \b{H}_{t-1}+ \b{b}_o)\\
\b{p}_t &= \textnormal{Tanh}(\b{W}_c \b{X}_t + \b{U}_c \b{H}_{t-1}+ \b{b}_c)\\
\b{c}_t &= (\b{f}_t \odot \b{c}_{t-1}) +  (\b{i}_t \odot \b{p}_t)\\
\b{h}_t &= \b{o}_t \odot \textnormal{Tanh}(\b{c}_{t})\\
\widehat{\b{Y}}_{t+h|t} &= \b{W}_y\b{h}_t +\b{b}_y
\end{split}
\]
where $\b{U}_f$, $\b{U}_i$, $\b{U}_o$ ,$\b{U}_c$ ,$\b{U}_f$, $\b{W}_f$, $\b{W}_i$, $\b{W}_o$, $\b{W}_c$, $\b{b}_f$, $\b{b}_i$, $\b{b}_o$, and $\b{b}_c$ are parameters to be estimated.
\end{enumerate}
\end{algorithm}

\subsection{Regression Trees}\label{S:RT}
\begin{figure}
\centering
\includegraphics[scale=0.65]{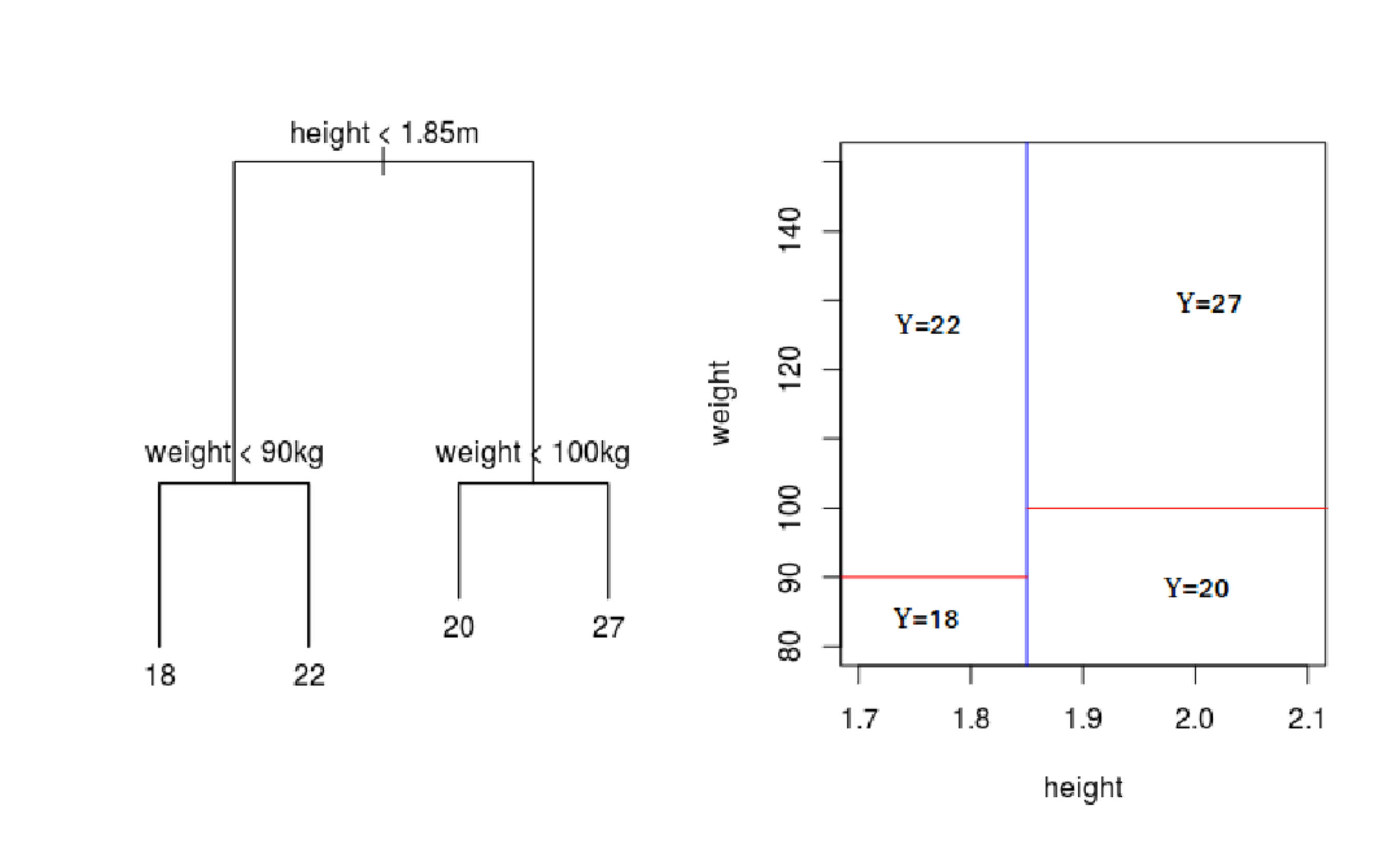}
\caption{Example of a simple tree.}
\label{F:exampletree}
\end{figure}

A regression tree is a nonparametric model that approximates an unknown nonlinear function $f_h(\b{X}_t)$ in \eqref{eq:model} with local predictions using recursive partitioning of the space of the covariates. A tree may be represented by a graph as in the left side of Figure \ref{F:exampletree}, which is equivalent as the partitioning in the right side of the figure for this bi-dimensional case. For example, suppose that we want to predict the scores of basketball players based on their height and weight. The first node of the tree in the example splits the players taller than 1.85m from the shorter players. The second node in the left takes the short players groups and split them by weights and the second node in the right does the same with the taller players. The prediction for each group is displayed in the terminal nodes and they are calculated as the average score in each group. To grow a tree we must find the optimal splitting point in each node, which consists of an optimal variable and an optimal observation. In the same example, the optimal variable in the first node is height and the observation is 1.85m.

The idea of regression trees is to approximate $f_h(\b{X}_t)$ by
\[
h_D(\b{X}_t) = \sum_{j=1}^{J_T} \beta_j I_j(\b{X}_t),\quad\textnormal{where}\quad
I_k(\b{X}_t)=
\begin{cases}
1 & \textnormal{if } \b{X}_t \in \mathcal{R}_j,\\
0 & \textnormal{otherwise}.
\end{cases}
\]
From the above expression, it becomes clear that the approximation of $f_h(\cdot)$ is equivalent to a linear regression on $J_T$ dummy variables, where $I_j(\b{X}_t)$ is a product of indicator functions.

\begin{figure}
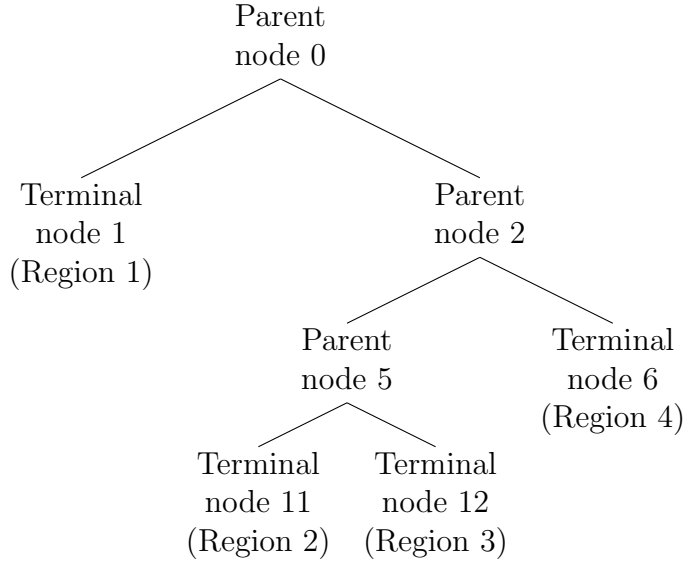

\[
\Tree [.{Parent \\ node 0} {Terminal \\ node 1\\ (Region 1)} [.{Parent \\ node 2} [.{Parent \\ node 5} {Terminal \\ node 11 \\ (Region 2)} {Terminal \\ node 12\\ (Region 3)} ] {Terminal \\ node 6\\ (Region 4)} ] ]
\]
\caption{Example of tree with labels.}
\label{F:example2}
\end{figure}

Let $J:=J_T$ and $N:=N_T$ be, respectively, the number of terminal nodes (regions, \emph{leaves}) and parent nodes. Different regions are denoted as $\mathcal{R}_1,\ldots,\mathcal{R}_J$. The root node at position $0$. The parent node at position $j$ has two split (child) nodes at positions $2j+1$ and $2j+2$. Each parent node has a threshold (split) variable associated, $X_{s_jt}$, where $s_j\in\mathbb{S}=\{1,2,\ldots,p\}$. Define $\mathbb{J}$ and $\mathbb{T}$ as the sets of parent and terminal nodes, respectively. Figure \ref{F:example2} gives an example. In the example, the parent nodes are $\mathbb{J}=\{0,2,5\}$ and the terminal nodes are $\mathbb{T}=\{1,6,11,12\}$.

Therefore, we can write the approximating model as
\begin{equation}
h_D(\b{X}_t)=\sum_{i\in\mathbb{T}}\beta_iB_{\mathbb{J}i}\left(\b{X}_t;\b{\theta}_i\right),
\end{equation}
where
\begin{equation}
B_{\mathbb{J}i}\left(\b{X}_t;\b{\theta}_i\right)=
\prod_{j\in\mathbb{J}}I(X_{s_j,t};c_j)^{\frac{n_{i,j}(1+n_{i,j})}{2}}
\times\left[
1-I(X_{s_j,t};c_j)
\right]^{(1-n_{i,j})(1+n_{i,j})},
\end{equation}
\[
I(X_{s_j,t};c_j)=
\begin{cases}
1 & \textnormal{if}\, X_{s_j,t}\leq c_j\\
0 & \textnormal{otherwise},
\end{cases}
\]
\[
n_{i,j}=
\begin{cases}
-1 & \text{if the path to leaf } \,i\, \text{does not include parent node } j; \\
0  & \text{if the path to leaf } \,i\, \text{include the \textbf{right-hand} child of parent node } j; \\
1  & \text{if the path to leaf } \,i\, \text{include the \textbf{left-hand} child of parent node } j. \\
\end{cases}
\]
$\mathbb{J}_i$: indexes of parent nodes included in the path to leaf $i$. $\b{\theta}_i=\{c_k\}$ such that $k\in\mathbb{J}_i$, $i\in\mathbb{T}$ and  $\sum_{j\in\mathbb{J}}B_{\mathbb{J}i}\left(\b{X}_t;\b{\theta}_j\right)=1$.

\subsubsection{Random Forests}
Random Forest (RF) is a collection of regression trees, each specified in a bootstrap sample of the original data. The method was originally proposed by \citet{breiman2001random}. Since we are dealing with time series, we use a block bootstrap. Suppose there are $B$ bootstrap samples. For each sample $b$, $b=1,\ldots,B$, a tree with $K_b$ regions is estimated for a randomly selected subset of the original regressors. $K_b$ is determined in order to leave a minimum number of observations in each region. The final forecast is the average of the forecasts of each tree applied to the original data:
\[
\widehat{Y}_{t+h|t}=\frac{1}{B}\sum_{b=1}^B\left[\sum_{i=1}^{\mathbb{T}_b} \widehat{\beta}_{i,b} B_{\mathbb{J}i,b}(\b{X}_t; \widehat{\b{\theta}}_{i,b})\right].
\]

The theory for RF models has been developed only to independent and identically distributed random variables. For instance, \citet{eSgBjV2015} proves consistency of the RF approximation to the unknown function $f_h(\b{X}_t)$. More recently, \citet{sWsA2018} proved consistency and asymptotic normality of the RF estimator.

\subsubsection{Boosting Regression Trees}

Boosting is another greedy method to approximate nonlinear functions that uses base learners for a sequential approximation. The model we consider here, called Gradient Boosting, was introduced by \citet{friedman2001greedy} and can be seen as a Gradient Descendent method in functional space.

The study of statistical properties of the Gradient Boosting is well developed for independent data. For example, for regression problems, \citet{duffy2002boosting} derived bounds on the convergence of boosting algorithms using assumptions on the performance of the base learner. \citet{zhang2005boosting} proves convergence, consistency and results on the speed of convergence with mild assumptions on the base learners. \citet{buhlmann2002consistency} shows similar results for consistency in the case of $\ell_2$ loss functions and three base models. Since boosting indefinitely leads to overfitting problems, some authors have demonstrated the consistency of boosting with different types of stopping rules, which are usually related to small step sizes, as suggested by \citet{friedman2001greedy}. Some of these works include boosting in classification problems and gradient boosting for both classification and regression problems. See, for instance, \citet{jiang2004process, lugosi2004bayes, bartlett2007adaboost, zhang2005boosting, buehlmann2006boosting, buhlmann2002consistency}.

Boosting is an iterative algorithm. The idea of boosted trees is to, at each iteration, sequentially refit the gradient of the loss function by small trees. In the case of quadratic loss as considered in this paper, the algorithm simply refit the residuals from the previous iteration.

Algorithm (\ref{A:2}) presents the simplified boosting procedure for a quadratic loss. It is recommended to use a shrinkage parameter $v\in (0,1]$ to control the learning rate of the algorithm. If $v$ is close to 1, we have a faster convergence rate and a better in-sample fit. However, we are more likely to have over-fitting and produce poor out-of-sample results. Additionally, the derivative is highly affected by over-fitting, even if we look at in-sample estimates. A learning rate between 0.1 and 0.2 is recommended to maintain a reasonable convergence ratio and to limit over-fitting problems.
\begin{algorithm}\label{A:2}
The boosting algorithm is defined as the following steps.
\begin{enumerate}
\item
Initialize $\phi_{i0} =\bar{Y}:=\frac{1}{T}\sum_{t=1}^TY_t$;
\item
For $m=1,\dots,M$:
\begin{enumerate}
\item
Make $U_{tm} = Y_t-\phi_{tm-1}$
\item
Grow a (small) Tree model to fit $u_{tm}$, $\widehat{u}_{tm} =  \sum_{i\in\mathbb{T}_m}\widehat{\beta}_{im} B_{\mathbb{J}_mi}(\b{X}_t;\widehat{\b{\theta}}_{im})$
\item
Make $\rho_m = \arg\underset{\rho}{\min} \sum_{t=1}^T[u_{tm}-\rho\widehat{u}_{tm}]^2$
\item
Update $\phi_{tm} = \phi_{tm-1}+v\rho_m\widehat{u}_{tm}$\;
 \end{enumerate}
\end{enumerate}
\end{algorithm}

The final fitted value may be written as
\begin{equation}
\begin{split}
    \widehat{Y}_{t+h} &= \bar{Y}+ \sum_{m=1}^M v\rho_m\widehat{u}_{tm}\\
     & =  \bar{Y}+ \sum_{m=1}^M v\widehat{\rho}_m \sum_{k\in\mathbb{T}_m}\widehat{\beta}_{km} B_{\mathbb{J}_mk}(\b{X}_t;\widehat{\b{\theta}}_{km})
\end{split}
\end{equation}

\subsection{Inference}

Conducting inference in nonlinear ML methods is tricky. One possible way is to follow \citet{mcMtTgR2006}, \citet{mcMaV2005} and \citet{sMcPmcM2004} and interpret particular nonlinear ML specifications as parametric models, as for example, general forms of smooth transition regressions. However, this approach restricts the application of ML methods to very specific settings. An alternative, is to consider models that can be cast in the sieves framework as described earlier. This is the case of splines and feed-forward NNs, for example. In this setup, \citet{xCsS1998} and \citet{cX2007} derived, under regularity conditions, the consistency and asymptotically normality of the estimates of a semi-parametric sieve approximations. Their setup is defined as follows:
\[
Y_{t+h} = \b\beta_0'\b X_t + f(\b X_t) + U_{t+h},
\]
where $f(\b X_t)$ is a nonlinear function that is nonparametrically modeled by sieve approximations. \citet{xCsS1998} and \citet{cX2007} consider both the estimation of the linear and nonlinear components of the model. However, their results are derived under the case where the dimension of $\b X_t$ is fixed.

Recently, \citet{double2017} and \citet{double2018} consider the case where the number of covariates diverge as the sample size increases in a very general setup. In this case the asymptotic results in \citet{xCsS1998} and \citet{cX2007} are not valid and the authors put forward the so-called double ML methods as a nice generalization to the results of \citet{aBvCcH2014a}. For Deep Neural Networks, \citet{mFtLsM2021} consider  semiparametric inference and establish nonasymptotic high probability bounds. Consequently, the authors are able to derive rates of convergence that are sufficiently fast to allow them to establish valid second‐step inference after first‐step estimation with deep learning. Nevertheless, the above papers do not include the case of time-series models.

More specifically to the case of Random Forests, asymptotic and inferential results are derived in \citet{eSgBjV2015} and \cite{sWsA2018} for the case of IID data. More recently, \citet{rDmN2020} prove a uniform concentration inequality for regression trees built on nonlinear autoregressive stochastic processes and prove consistency for a large class of random forests. Finally, it is worth mentioning the interesting work of \citet{dBbCnAmN2020}. In their paper, the authors show that proper predictor targeting controls the probability of placing splits along strong predictors and improves prediction.

\section{Other Methods}\label{S:Other}

\subsection{Bagging}\label{S:bagging}

The term \emph{bagging} means \emph{Bootstrap Aggregating} and was proposed by \cite{lB1996} to reduce the variance of unstable predictors\footnote{An \emph{unstable predictor} has large variance. Intuitively, small changes in the data yield large changes in the predictive model}. It was popularized in the time series literature by \citet{aIlK2008}, who to construct forecasts from multiple regression models with local-to-zero regression parameters and errors subject to possible serial correlation or conditional heteroscedasticity. Bagging is designed for situations in which the number of predictors is moderately large relative to the sample size.

The bagging algorithm in time series settings have to take into account the time dependence dimension when constructing the bootstrap samples.
\begin{algorithm}[Bagging for Time-Series Models]
\label{alg:bagging}
The Bagging algorithm is defined as follows.
\begin{enumerate}
\item[1.]
Arrange the set of tuples $\left(y_{t+h},\b{x}'_t\right)$, $t=h+1,\ldots,T$, in the form of a matrix $\b{V}$ of dimension $(T-h)\times n$.
\item[2.]
Construct (block) bootstrap samples of the form $\left\{\left(y_{(i)2}^\ast,\b{x}'^*_{(i)2}\right),\ldots,\left(y^{*}_{(i)T},\b{x}'^*_{(i)T}\right)\right\}$, $i=1,\ldots,B$, by drawing blocks of $M$ rows of $\b{V}$ with replacement.
\item[3.]
Compute the $i$th bootstrap forecast as
\begin{equation}
\widehat{y}^\ast_{(i)t+h|t} =
\begin{cases}
0 & \text{if } |t^*_j| < c \,\forall j,\\
\widehat{\boldsymbol{\lambda}}_{(i)}^*\widetilde{\b{x}}^*_{(i)t}
& \text{otherwise},
\end{cases}
\end{equation}
where
$\widetilde{\b{x}}^*_{(i)t}:=\b{S}_{(i)t}^*\b{z}_{(i)t}^*$ and $\b{S}_t$ is a diagonal selection matrix with $j$th
diagonal element given by
\[
\mathbb{I}_{\{|t_j|>c\}}=
\begin{cases}
1 & \text{if }|t_j|>c,\\
0 & \text{otherwise,}
\end{cases}
\]
$c$ is a pre-specified critical value of the test. $\widehat{\boldsymbol{\lambda}}_{(i)}^*$ is the OLS estimator at each bootstrap repetition.
\item[4.]
Compute the average forecasts over the bootstrap samples:
\[
\tilde{y}_{t+h|t} =
\frac{1}{B}\sum_{i=1}^B\widehat{y}^\ast_{(i)t|t-1}.
\]
\end{enumerate}
\end{algorithm}

In algorithm \ref{alg:bagging}, above, one requires that it is possible to estimate and conduct inference in the linear model. This is certainly infeasible if the number of predictors is larger than the sample size ($n>T$), which requires the algorithm to be modified. \citet{mGmMgV2017} and \citet{mcMgVaVeZ2019} adopt the following changes of the algorithm:
\begin{algorithm}[Bagging for Time-Series Models and Many Regressors]
The Bagging algorithm is defined as follows.
\begin{enumerate}
\item[0.]
Run $n$ univariate regressions of $y_{t+h}$ on each covariate in $\b{x}_t$. Compute $t$-statistics and keep only the ones that turn out to be significant at a given pre-specified level. Call this new set of regressors as $\check{\b{x}}_t$
\item[1--4.]
Same as before but with $\b{x}_t$ replaced by $\check{\b{x}}_t$.
\end{enumerate}
\end{algorithm}

\subsection{Complete Subset Regression}\label{S:CSR}
Complete Subset Regression (CSR) is a method for combining forecasts developed by \citet{gEaGaT2013,gEaGaT2015}. The motivation was that selecting the optimal subset of $\b{X}_t$ to predict $Y_{t+h}$ by testing all possible combinations of regressors is computationally very demanding and, in most cases, unfeasible. For a given set of potential predictor variables, the idea is to combine forecasts by averaging\footnote{It is possible to combine forecasts using any weighting scheme. However, it is difficult to beat uniform weighting \cite{vGgKaMaT2013}.} all possible linear regression models with fixed number of predictors. For example, with $n$ possible predictors, there are $n$ unique univariate models and
\[
n_{k,n}=\frac{n!}{(n-k)!k!}
\]
different $k$-variate models for $k\leq K$. The set of models for a fixed value of $k$ as is known as the complete subset.

When the set of regressors is large the number of models to be estimated increases rapidly. Moreover, it is likely that many potential predictors are irrelevant. In these cases it was suggested that one should include only a small, $k$, fixed set of predictors, such as five or ten. Nevertheless, the number of models still very large, for example,  with $n=30$ and $k=8$, there are $5,852,925$ regression. An alternative solution is to follow \citet{mGmMgV2017} and \citet{mcMgVaVeZ2019} and adopt a similar strategy as in the case of Bagging high-dimensional models. The idea is to start fitting a regression of $Y_{t+h}$ on each of the candidate variables and save the $t$-statistics of each variable. The $t$-statistics are ranked by absolute value, and we select the $\tilde{n}$ variables that are more relevant in the ranking. The CSR forecast is calculated on these variables for different values of $k$. This approach is based on the the Sure Independence Screening of \cite{fanlv2008}, extended to dependent  by \cite{yousuf2018}, that aims to select a superset of relevant predictors among a very large set.

\subsection{Hybrid Methods}

Recently, \citet{mMeM2013} proposed the combination of LASSO-based estimation and NN models. The idea is to construct a feedforward single-hidden layer NN where the parameters of the nonlinear terms (neurons) are randomly generated and the linear parameters are estimated by LASSO (or one of its generalizations). Similar ideas were also considered by \citet{aKtT2014} and \citet{aKtT2016}.

\citet{aTfLkH2000} and \citet{mcMtTgR2006} proposed to augment a feedforward shallow NN by a linear term. The motivation is that the nonlinear component should capture only the nonlinear dependence, making the model more interpretable. This is in the same spirit of the semi-parametric models considered in \citet{cX2007}.

Inspired by the above ideas, \citet{mcMgVaVeZ2019} proposed combining random forests with adaLASSO and OLS. The authors considered two specifications. In the first one, called RF/OLS, the idea is to use the variables selected by a Random Forest in a OLS regression. The second approach, named adaLASSO/RF, works in the opposite direction. First select the variables by adaLASSO and than use them in a Random Forest model. The goal is to disentangle the relative importance of variable selection and nonlinearity to forecast inflation.

Recently, \citet{fDmS2019} propose the ``partially-egalitarian'' LASSO to combine survey forecasts. More specifically, the procedure sets some combining weights to zero and shrinks the survivors toward equality. Therefore, the final forecast will be close related to the simple average combination of the survived forecasts. Although the paper considers survey forecasts, the method is quite general and can be applied to any set of forecasts. As pointed out by the authors, optimally-regularized regression-based combinations and subset-average combinations are very closely connected. \citet{fDmSbZ2021} extended the ideas in \citet{fDmS2019} in order to construct regularized mixtures of density forecasts. Both papers shed light on how machine learning methods can be used to optimally combine a large set of forecasts.

\section{Forecast Comparison}\label{S:Comp}

With the advances in the ML literature, the number of available forecasting models and methods have been increasing at a fast pace. Consequently, it is very important to apply statistical tools to compare different models. The forecasting literature provides a number of tests since the seminal paper by \citet{fxDrsM1995} that can be applied as well to the ML models described in this survey.

In the Diebold and Mariano’s (1995) test, two competing methods have the same unconditional expected loss under the null hypothesis, and the test can be carried out using a simple t-test. A small sample adjustment was developed by \citet{dHsLpN1997}. See also the recent discussion in \citet{fD2015}. One drawback of the  Diebold and Mariano’s (1995) test is that its statistic diverges under null when the competing models are nested. However, \citet{rGhW2006} show that the test is valid if the forecasts are derived from models estimated in a rolling window framework. Recently, \citet{mM2020} shows that if the estimation window is fixed, the Diebold and Mariano’s (1995) statistic may diverge under the null. Therefore, it is very important that the forecasts are computed in a rolling window scheme.

In order to accommodate cases where there are more than two competing models, an unconditional superior predictive ability (USPA) test was proposed by \citet{hW2000}. The null hypothesis states that a benchmark method outperforms a set of competing alternatives. However, \citet{pR2005} showed that White's (2000) test can be very conservative when there are competing methods that are inferior to the benchmark. Another important contribution to the forecasting literature is the model confidence set (MCS) proposed by \citet{pHaLjN2011}. A MCS is a set of competing models that is built in a way to contain the best model with respect to a certain loss function and with a given level of confidence. The MCS acknowledges the potential limitations of the dataset, such that uninformative data yield a MCS with a large number models, whereas informative data yield a MCS with only a few models. Importantly, the MCS procedure does not assume that a particular model is the true one.

Another extension of the Diebold and Mariano's (1995) test is the conditional equal predictive ability (CEPA) test proposed by \citet{rGhW2006}. In practical applications, it is important to know not only if a given model is superior but also when it is better than the alternatives. Recently, \citet{jLzLrQ2020} proposed a very general framework to conduct conditional predictive ability tests.

In summary, it is very important to compare the forecasts from different ML methods and the literature provides a number of tests that can be used.

\section{Applications of Machine Learning Methods to Economic and Financial Forecasting}\label{S:App}

\subsection{Linear Methods}

Penalized regressions are now an important option in the toolkit of applied economists are there is a vast literature considering the use of such techniques to economics and financial forecasting.

Macroeconomic forecasting is certainly one of the most successful applications of penalized regressions. \citet{mMeM2016} applied the adaLASSO to forecasting US inflation and showed that the method outperforms the linear autoregressive and factor models. \citet{mcMgV2016} show that high-dimensional linear models produce, on average, smaller forecasting errors for macroeconomic variables when a large set of predictors is considered. Their results also indicate that a good selection of the adaLASSO hyperparameters reduces forecasting errors. \citet{mGmMgV2017} show that high-dimensional econometric models, such as shrinkage and complete subset regression, perform very well in real time forecasting of Brazilian inflation in data-rich environments. The authors combine forecasts of different alternatives and show that model combination can achieve superior predictive performance. \citet{sSeW2018} consider an application to a large macroeconomic US dataset and demonstrate that penalized regressions are very competitive. \citet{mcMgVaVeZ2019} conduct a vast comparison of models to forecast US inflation and showed the penalized regressions were far superior than several benchmarks, including factor models. \citet{dAkBkB2019} introduce a general text sentiment framework that optimizes the design for forecasting purposes and apply it to forecasting economic growth in the US. The method includes the use of the elastic net for sparse data-driven selection and the weighting of thousands of sentiment values. \citet{aT2019} consider penalized VARs to forecast six different economic uncertainty variables for the growth of the real M2 and real M4 Divisia money series for the US using monthly data. \citet{yUsT2019} consider high-dimensional forecasting and variable selection via folded-concave penalized regressions. The authors forecast quarterly US gross domestic product data using a high-dimensional monthly data set and the mixed data sampling (MIDAS) framework with penalization. See also \citet{aBeGjS2020a} and \citet{aBeGjS2020b}.

There is also a vast list of applications in empirical finance. \citet{gEaGaT2013} find that combinations of subset regressions can produce more accurate forecasts of the equity premium than conventional approaches based on equal-weighted forecasts and other regularization techniques. \citet{fAsK2016} used LASSO-based methods to estimated forecasting models for realized volatilities. \citet{lCaKmM2017} consider modelling and forecasting large realized covariance matrices of the 30 Dow Jones stocks by penalized vector autoregressive (VAR) models. The authors find that penalized VARs outperform the benchmarks by a wide margin and improve the portfolio construction of a mean-variance investor.  \citet{aCaCmY2019} use the LASSO to make 1-minute-ahead return forecasts for a vast set of stocks traded at the New York Stock Exchange. The authors provide evidence that penalized regression estimated by the LASSO boost out-of-sample predictive power by choosing predictors that trace out the consequences of unexpected news announcements.

\subsection{Nonlinear Methods}

There are many papers on the application of nonlinear ML methods to economic and financial forecasting. Most of the papers focus on NN methods, specially the ones from the early literature.

With respect to the early papers, most of the models considered were nonlinear versions of autoregressive models. At best, a small number of extra covariates were included. See, for example, \citet{tTdDmM2005} and the references therein. In the majority of the papers, including \citet{tTdDmM2005}, there was no strong evidence of the superiority of nonlinear models as the differences in performance were marginal. Other examples from the early literature are \citet{nrShW1995}, \citet{nrShW1997a}, \citet{nrShW1997b}, \citet{sdBjkO2000}, \citet{gT2001}, \citet{mcMaVcP2001}, and \citet{sHdOcB2004}.

More recently, with the availability of large datasets, nonlinear models are back to the scene. For example, \citet{mcMgVaVeZ2019} show that, despite the skepticism of the previous literature on inflation forecasting, ML models with a large number of covariates are systematically more accurate than the benchmarks for several forecasting horizons and show that Random Forests dominated all other models. The good performance of the Random Forest is due not only to its specific method of variable selection but also the potential nonlinearities between past key macroeconomic variables and inflation. Other successful example is \citet{sGbKdX2020}. The authors show large economic gains to investors using ML forecasts of future stock returns based on a very large set of predictors. The best performing models are tree-based and neural networks. \citet{coulombe2020} show significant gains when nonlinear ML methods are used to forecast macroeconomic time series. \citet{dBeC2020} consider penalized regressions, ensemble methods, and random forest to forecast employment growth in the United States over the period 2004–2019 using Google search activity. Their results strongly indicate that Google search data have predictive power. \citet{aBdReS2020} compute now- and backcasts of weekly unemployment insurance initial claims in the US based on a rich set of daily Google Trends search-volume data and machine learning methods.

\subsection{Empirical Illustration}\label{S:Empirical}

In this section we illustrate the use of some of the methods reviewed in this paper to forecast daily realized variance of the Brazilian Stock Market index (BOVESPA). We use as regressors information from other major indexes, namely, the S\&P500 (US), the FTSE100 (United Kingdom), DAX (Germany), Hang Seng (Hong Kong), and Nikkei (Japan). Our measure of realized volatility is constructed by aggregating intraday returns sample at the 5-minute frequency. The data were obtained from the Oxford-Man Realized Library at Oxford University.\footnote{https://realized.oxford-man.ox.ac.uk/data/assets}

For each stock index, we define the realized variance as
\[
RV_t=\sum_{s=1}^{S}r^2_{st},
\]
where $r_{st}$ is the log return sampled at the five-minute frequency. $S$ is the number of available returns at day $t$.

The benchmark model is the Heterogeneous Autoregressive (HAR) model proposed by \citet{fC2009}:
\begin{equation}\label{E:HAR}
\log RV_{t+1}=\beta_0 + \beta_1\log RV_{t} + \beta_5\log RV_{5,t} + \beta_22\log RV_{22,t} + U_{t+1},
\end{equation}
where $RV_t$ is daily realized variance of the BOVESPA index,
\[
\begin{split}
RV_{5,t}&=\frac{1}{5}\sum_{i=0}^4RV_{t-i},\quad\textnormal{and}\\
RV_{22,t}&=\frac{1}{22}\sum_{i=0}^{21}RV_{t-i}.
\end{split}
\]

As alternatives we consider a extended HAR model with additional regressors estimated by adaLASSO. We include as extra regressors the daily past volatility of the other five indexes considered here. The model has a total of eight candidate predictors. Furthermore, we consider two nonlinear alternatives using all predictors: a random forest and shallow and deep neural networks.

\begin{figure}
    \centering
    \includegraphics[scale=1]{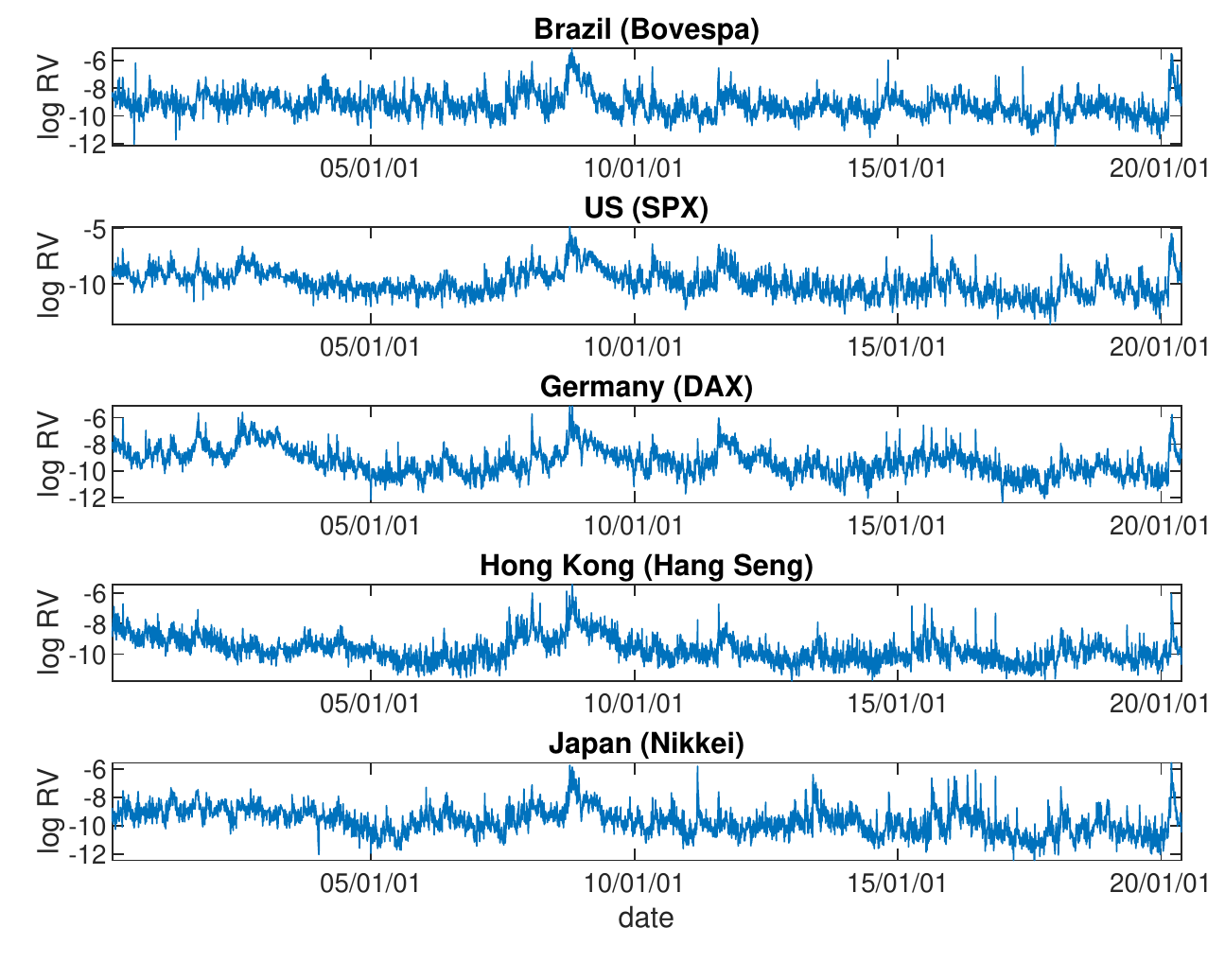}
    \caption{Realized variance of different stock indexes}
    \label{F:RV}
\end{figure}

The realized variances of the different indexes are illustrated in Figure \ref{F:RV}. The data starts in February 2, 2000 and ends in May 21, 2020, a total of 4,200 observations. The sample includes two periods of very high volatility, namely the financial crisis of 2007-2008 and the Covid-19 pandemics of 2020. We consider a rolling window exercise, were we set 1,500 observations in each window. The models are re-estimated every day.

Several other authors have estimated nonlinear and machine learning models to forecast realized variances. \citet{mMmcM2008} considered a smooth transition version of the HAR while \citet{eHmM2016} considered the combination of smooth transitions, long memory and neural network models. \citet{eHmM2010} and \citet{mMmcM2011} combined NN models with bagging and \citet{mSmM2009} considered smooth transition regression trees. The use of LASSO and its generalizations to estimate extensions of the HAR model was proposed by \citet{fAsK2016}.

Although the models are estimated in logarithms, we report the results in levels, which in the end is the quantity of interest. We compare the models according to the Mean Squared Error (MSE) and the QLIKE metric.

The results are shown in Table \ref{T:RV}. The table reports for each model, the mean squared error (MSE) and the QLIKE statistics as a ratio to the HAR benchmark. Values smaller than one indicates that the model outperforms the HAR. The asterisks indicate the results of the Diebold-Mariano test of equal forecasting performance. *,**, and ***, indicate rejection of the null of equal forecasting ability at the 10\%, 5\% and 1\%, respectively. We report results for the full out-of-sample period, the financial crisis years (2007-2008) and the for 2020 as a way to capture the effects of the Covid-19 pandemics on the forecasting performance of different models.

As we can see from the tables the ML methods considered here outperform the HAR benchmark. The winner model is definitely the HAR model with additional regressors and estimated with adaLASSO. The performance improves during the high volatility periods and the gains reach 10\% during the Covid-19 pandemics. Random Forests do not perform well. On the other hand NN models with different number of hidden layers outperform the benchmark.

\begin{table}
\centering
\caption{Forecasting Results}
\label{T:RV}
\begin{minipage}{1\linewidth}
\begin{footnotesize}
The table reports for each model, the mean squared error (MSE) and the QLIKE statistics as a ratio to the HAR benchmark. Values smaller than one indicates that the model outperforms the HAR. The asterisks indicate the results of the Diebold-Mariano test of equal forecasting performance. *,**, and ***, indicate rejection of the null of equal forecasting ability at the 10\%, 5\% and 1\%, respectively.
\end{footnotesize}
\end{minipage}
\begin{threeparttable}
\begin{tabular}{lccccccccc}
\hline
&& \multicolumn{2}{c}{\textbf{Full Sample}} && \multicolumn{2}{c}{\textbf{2007-2008}}&&\multicolumn{2}{c}{\textbf{2020}}\\
\cline{3-4}
\cline{6-7}
\cline{9-10}
\multicolumn{1}{c}{\textbf{Model}} && \textbf{MSE} & \textbf{QLIKE} && \textbf{MSE} & \textbf{QLIKE} && \textbf{MSE} & \textbf{QLIKE} \\
\hline
HARX-LASSO         &&  $0.96^{**}$ & $0.98$ &&  $0.98^{*}$ & $0.96$ && $0.90^{***}$ & $0.90$\\
Random Forest      &&  $1.00$ & $1.02$ &&  $0.95^{***}$ & $0.98$ && $1.13^{**}$ & $1.03^{*}$ \\
Neural Network (1) &&  $0.99^{**}$ & $0.99$ &&  $0.97^{**}$ & $0.98$ && $0.99$ & $0.99$\\
Neural Network (3) &&   $0.99^{**}$ & $0.99$ &&  $0.98^{*}$ & $0.99$ && $0.99$ & $0.99$\\
Neural Network (5) &&   $0.90^{**}$ & $0.99$ &&  $0.98^{*}$ & $0.99$ && $0.99$ & $0.99$\\
\hline
\end{tabular}
\end{threeparttable}
\end{table}

\section{Conclusions and the Road Ahead}\label{S:Conclusions}
In this paper we present a non-exhaustive review of the most of the recent developments in machine learning and high-dimensional statistics to time-series modeling and forecasting. We presented both linear and nonlinear alternatives. Furthermore, we consider ensemble and hybrid models. Finally, we briefly discuss tests for superior predictive ability.

Among linear specification, we pay special attention to penalized regression (Ridge, LASSO and its generalizations, for example) and ensemble methods (Bagging and Complete Subset Regression). Although, there has been major theoretical advances in the literature on penalized linear regression models for dependent data, the same is not true for ensemble methods. The theoretical results for Bagging are so far based on independent data and the results for complete subset regression are quite limited.

With respect to nonlinear ML methods, we focused on neural networks and tree-based methods. Theoretical results for random forests and boosted trees have been developed only to independent and identically distributed data and in the case of a low dimensional set of regressors. For shallow neural networks, \citet{xCjRnS2001} and \citet{cX2007} provide some theoretical results for dependent data in the low dimensional case. The behavior of such models in high-dimensions is still under study. The same is true for deep neural networks.

Nevertheless, the recent empirical evidence shows that nonlinear machine learning models combined with large datasets can be extremely useful for economic forecasting.

As a direction for further developments we list the following points:
\begin{enumerate}
\item
Develop results for Bagging and Boosting for dependent data.
\item
Show consistency and asymptotic normality of the random forecast estimator of the unknown function $f_h(\b{X}_t)$ when the data are dependent.
\item
Derive a better understanding of the variable selection mechanism of nonlinear ML methods.
\item
Develop inferential methods to access variable importance in nonlinear ML methods.
\item
Develop models based on unstructured data, such as text data, to economic forecasting.
\item
Evaluate ML models for nowcasting.
\item
Evaluate ML in very unstable environments with many structural breaks.
\end{enumerate}

Finally, we would like to point that we left a number of other interesting ML methods out of this survey, such as, for example, Support Vector Regressions, autoenconders, nonlinear factor models, and many more. However, we hope that the material presented here can be of value to anyone interested of applying ML techniques to economic and/or financial forecasting.

\end{document}